\documentstyle[multicol,pre,aps,subeqnarray]{revtex}

\newcommand{\beq}{\begin{equation}}
\newcommand{\eeq}{\end{equation}}
\newcommand{\mod}{\hbox{ mod }}
\newcommand{\sign}{\hbox{sign}\,}

\begin{document}
% Psfig/TeX 
\def\PsfigVersion{1.9}
\ifx\undefined\psfig\else \fi

%
% from a suggestion by eijkhout@csrd.uiuc.edu to allow
% loading as a style file. Changed to avoid problems
% with amstex per suggestion by jbence@math.ucla.edu

\let\LaTeXAtSign=\@
\let\@=\relax
\edef\psfigRestoreAt{\catcode`\@=\number\catcode`@\relax}
\catcode`\@=11\relax
\newwrite\@unused
\def\ps@typeout#1{{\let\protect\string\immediate\write\@unused{#1}}}
\ps@typeout{psfig/tex \PsfigVersion}

%% Here's how you define your figure path.  Should be set up with null
%% default and a user useable definition.

\def\figurepath{./}
\def\psfigurepath#1{\edef\figurepath{#1}}

%
% @psdo control structure -- similar to Latex @for.
% I redefined these with different names so that psfig can
% be used with TeX as well as LaTeX, and so that it will not 
% be vunerable to future changes in LaTeX's internal
% control structure,
%
\def\@nnil{\@nil}
\def\@empty{}
\def\@psdonoop#1\@@#2#3{}
\def\@psdo#1:=#2\do#3{\edef\@psdotmp{#2}\ifx\@psdotmp\@empty \else
    \expandafter\@psdoloop#2,\@nil,\@nil\@@#1{#3}\fi}
\def\@psdoloop#1,#2,#3\@@#4#5{\def#4{#1}\ifx #4\@nnil \else
       #5\def#4{#2}\ifx #4\@nnil \else#5\@ipsdoloop #3\@@#4{#5}\fi\fi}
\def\@ipsdoloop#1,#2\@@#3#4{\def#3{#1}\ifx #3\@nnil 
       \let\@nextwhile=\@psdonoop \else
      #4\relax\let\@nextwhile=\@ipsdoloop\fi\@nextwhile#2\@@#3{#4}}
\def\@tpsdo#1:=#2\do#3{\xdef\@psdotmp{#2}\ifx\@psdotmp\@empty \else
    \@tpsdoloop#2\@nil\@nil\@@#1{#3}\fi}
\def\@tpsdoloop#1#2\@@#3#4{\def#3{#1}\ifx #3\@nnil 
       \let\@nextwhile=\@psdonoop \else
      #4\relax\let\@nextwhile=\@tpsdoloop\fi\@nextwhile#2\@@#3{#4}}
% 
% \fbox is defined in latex.tex; so if \fbox is undefined, assume that
% we are not in LaTeX.
% Perhaps this could be done better???
\ifx\undefined\fbox
% \fbox code from modified slightly from LaTeX
\newdimen\fboxrule
\newdimen\fboxsep
\newdimen\ps@tempdima
\newbox\ps@tempboxa
\fboxsep = 3pt
\fboxrule = .4pt
\long\def\fbox#1{\leavevmode\setbox\ps@tempboxa\hbox{#1}\ps@tempdima\fboxrule
    \advance\ps@tempdima \fboxsep \advance\ps@tempdima \dp\ps@tempboxa
   \hbox{\lower \ps@tempdima\hbox
  {\vbox{\hrule height \fboxrule
          \hbox{\vrule width \fboxrule \hskip\fboxsep
          \vbox{\vskip\fboxsep \box\ps@tempboxa\vskip\fboxsep}\hskip 
                 \fboxsep\vrule width \fboxrule}
                 \hrule height \fboxrule}}}}
\fi
%
%%%%%%%%%%%%%%%%%%%%%%%%%%%%%%%%%%%%%%%%%%%%%%%%%%%%%%%%%%%%%%%%%%%
% file reading stuff from epsf.tex
%   EPSF.TEX macro file:
%   Written by Tomas Rokicki of Radical Eye Software, 29 Mar 1989.
%   Revised by Don Knuth, 3 Jan 1990.
%   Revised by Tomas Rokicki to accept bounding boxes with no
%      space after the colon, 18 Jul 1990.
%   Portions modified/removed for use in PSFIG package by
%      J. Daniel Smith, 9 October 1990.
%
\newread\ps@stream
\newif\ifnot@eof       % continue looking for the bounding box?
\newif\if@noisy        % report what you're making?
\newif\if@atend        % %%BoundingBox: has (at end) specification
\newif\if@psfile       % does this look like a PostScript file?
%
% PostScript files should start with `%!'
%
{\catcode`\%=12\global\gdef\epsf@start{%!}}
\def\epsf@PS{PS}
\def\epsf@getbb#1{%
%
%   The first thing we need to do is to open the
%   PostScript file, if possible.
%
\openin\ps@stream=#1
\ifeof\ps@stream\ps@typeout{Error, File #1 not found}\else
%
%   Okay, we got it. Now we'll scan lines until we find one that doesn't
%   start with %. We're looking for the bounding box comment.
%
   {\not@eoftrue \chardef\other=12
    \def\do##1{\catcode`##1=\other}\dospecials \catcode`\ =10
    \loop
       \if@psfile
	  \read\ps@stream to \epsf@fileline
       \else{
	  \obeyspaces
          \read\ps@stream to \epsf@tmp\global\let\epsf@fileline\epsf@tmp}
       \fi
       \ifeof\ps@stream\not@eoffalse\else
%
%   Check the first line for `%!'.  Issue a warning message if its not
%   there, since the file might not be a PostScript file.
%
       \if@psfile\else
       \expandafter\epsf@test\epsf@fileline:. \\%
       \fi
%
%   We check to see if the first character is a % sign;
%   if so, we look further and stop only if the line begins with
%   `%%BoundingBox:' and the `(atend)' specification was not found.
%   That is, the only way to stop is when the end of file is reached,
%   or a `%%BoundingBox: llx lly urx ury' line is found.
%
          \expandafter\epsf@aux\epsf@fileline:. \\%
       \fi
   \ifnot@eof\repeat
   }\closein\ps@stream\fi}%
%
% This tests if the file we are reading looks like a PostScript file.
%
\long\def\epsf@test#1#2#3:#4\\{\def\epsf@testit{#1#2}
			\ifx\epsf@testit\epsf@start\else
\ps@typeout{Warning! File does not start with `\epsf@start'.  It may not be a PostScript file.}
			\fi
			\@psfiletrue} % don't test after 1st line
%
%   We still need to define the tricky \epsf@aux macro. This requires
%   a couple of magic constants for comparison purposes.
%
{\catcode`\%=12\global\let\epsf@percent=%\global\def\epsf@bblit{%BoundingBox}}
%
%
%   So we're ready to check for `%BoundingBox:' and to grab the
%   values if they are found.  We continue searching if `(at end)'
%   was found after the `%BoundingBox:'.
%
\long\def\epsf@aux#1#2:#3\\{\ifx#1\epsf@percent
   \def\epsf@testit{#2}\ifx\epsf@testit\epsf@bblit
	\@atendfalse
        \epsf@atend #3 . \\%
	\if@atend	
	   \if@verbose{
		\ps@typeout{psfig: found `(atend)'; continuing search}
	   }\fi
        \else
        \epsf@grab #3 . . . \\%
        \not@eoffalse
        \global\no@bbfalse
        \fi
   \fi\fi}%
%
%   Here we grab the values and stuff them in the appropriate definitions.
%
\def\epsf@grab #1 #2 #3 #4 #5\\{%
   \global\def\epsf@llx{#1}\ifx\epsf@llx\empty
      \epsf@grab #2 #3 #4 #5 .\\\else
   \global\def\epsf@lly{#2}%
   \global\def\epsf@urx{#3}\global\def\epsf@ury{#4}\fi}%
%
% Determine if the stuff following the %%BoundingBox is `(atend)'
% J. Daniel Smith.  Copied from \epsf@grab above.
%
\def\epsf@atendlit{(atend)} 
\def\epsf@atend #1 #2 #3\\{%
   \def\epsf@tmp{#1}\ifx\epsf@tmp\empty
      \epsf@atend #2 #3 .\\\else
   \ifx\epsf@tmp\epsf@atendlit\@atendtrue\fi\fi}

% End of file reading stuff from epsf.tex
%%%%%%%%%%%%%%%%%%%%%%%%%%%%%%%%%%%%%%%%%%%%%%%%%%%%%%%%%%%%%%%%%%%

%%%%%%%%%%%%%%%%%%%%%%%%%%%%%%%%%%%%%%%%%%%%%%%%%%%%%%%%%%%%%%%%%%%
% trigonometry stuff from "trig.tex"
\chardef\psletter = 11 % won't conflict with \begin{letter} now...
\chardef\other = 12

\newif \ifdebug %%% turn me on to see TeX hard at work ...
\newif\ifc@mpute %%% don't need to compute some values
\c@mputetrue % but assume that we do

\let\then = \relax
\def\r@dian{pt }
\let\r@dians = \r@dian
\let\dimensionless@nit = \r@dian
\let\dimensionless@nits = \dimensionless@nit
\def\internal@nit{sp }
\let\internal@nits = \internal@nit
\newif\ifstillc@nverging
\def \Mess@ge #1{\ifdebug \then \message {#1} \fi}

{ %%% Things that need abnormal catcodes %%%
	\catcode `\@ = \psletter
	\gdef \nodimen {\expandafter \n@dimen \the \dimen}
	\gdef \term #1 #2 #3%
	       {\edef \t@ {\the #1}%%% freeze parameter 1 (count, by value)
		\edef \t@@ {\expandafter \n@dimen \the #2\r@dian}%
				   %%% freeze parameter 2 (dimen, by value)
		\t@rm {\t@} {\t@@} {#3}%
	       }
	\gdef \t@rm #1 #2 #3%
	       {{%
		\count 0 = 0
		\dimen 0 = 1 \dimensionless@nit
		\dimen 2 = #2\relax
		\Mess@ge {Calculating term #1 of \nodimen 2}%
		\loop
		\ifnum	\count 0 < #1
		\then	\advance \count 0 by 1
			\Mess@ge {Iteration \the \count 0 \space}%
			\Multiply \dimen 0 by {\dimen 2}%
			\Mess@ge {After multiplication, term = \nodimen 0}%
			\Divide \dimen 0 by {\count 0}%
			\Mess@ge {After division, term = \nodimen 0}%
		\repeat
		\Mess@ge {Final value for term #1 of 
				\nodimen 2 \space is \nodimen 0}%
		\xdef \Term {#3 = \nodimen 0 \r@dians}%
		\aftergroup \Term
	       }}
	\catcode `\p = \other
	\catcode `\t = \other
	\gdef \n@dimen #1pt{#1} %%% throw away the ``pt''
}

\def \Divide #1by #2{\divide #1 by #2} %%% just a synonym

\def \Multiply #1by #2%%% allows division of a dimen by a dimen
       {{%%% should really freeze parameter 2 (dimen, passed by value)
	\count 0 = #1\relax
	\count 2 = #2\relax
	\count 4 = 65536
	\Mess@ge {Before scaling, count 0 = \the \count 0 \space and
			count 2 = \the \count 2}%
	\ifnum	\count 0 > 32767 %%% do our best to avoid overflow
	\then	\divide \count 0 by 4
		\divide \count 4 by 4
	\else	\ifnum	\count 0 < -32767
		\then	\divide \count 0 by 4
			\divide \count 4 by 4
		\else
		\fi
	\fi
	\ifnum	\count 2 > 32767 %%% while retaining reasonable accuracy
	\then	\divide \count 2 by 4
		\divide \count 4 by 4
	\else	\ifnum	\count 2 < -32767
		\then	\divide \count 2 by 4
			\divide \count 4 by 4
		\else
		\fi
	\fi
	\multiply \count 0 by \count 2
	\divide \count 0 by \count 4
	\xdef \product {#1 = \the \count 0 \internal@nits}%
	\aftergroup \product
       }}

\def\r@duce{\ifdim\dimen0 > 90\r@dian \then   % sin(x+90) = sin(180-x)
		\multiply\dimen0 by -1
		\advance\dimen0 by 180\r@dian
		\r@duce
	    \else \ifdim\dimen0 < -90\r@dian \then  % sin(-x) = sin(360+x)
		\advance\dimen0 by 360\r@dian
		\r@duce
		\fi
	    \fi}

\def\Sine#1%
       {{%
	\dimen 0 = #1 \r@dian
	\r@duce
	\ifdim\dimen0 = -90\r@dian \then
	   \dimen4 = -1\r@dian
	   \c@mputefalse
	\fi
	\ifdim\dimen0 = 90\r@dian \then
	   \dimen4 = 1\r@dian
	   \c@mputefalse
	\fi
	\ifdim\dimen0 = 0\r@dian \then
	   \dimen4 = 0\r@dian
	   \c@mputefalse
	\fi
	\ifc@mpute \then
        	% convert degrees to radians
		\divide\dimen0 by 180
		\dimen0=3.141592654\dimen0
		\dimen 2 = 3.1415926535897963\r@dian %%% a well-known constant
		\divide\dimen 2 by 2 %%% we only deal with -pi/2 : pi/2
		\Mess@ge {Sin: calculating Sin of \nodimen 0}%
		\count 0 = 1 %%% see power-series expansion for sine
		\dimen 2 = 1 \r@dian %%% ditto
		\dimen 4 = 0 \r@dian %%% ditto
		\loop
			\ifnum	\dimen 2 = 0 %%% then we've done
			\then	\stillc@nvergingfalse 
			\else	\stillc@nvergingtrue
			\fi
			\ifstillc@nverging %%% then calculate next term
			\then	\term {\count 0} {\dimen 0} {\dimen 2}%
				\advance \count 0 by 2
				\count 2 = \count 0
				\divide \count 2 by 2
				\ifodd	\count 2 %%% signs alternate
				\then	\advance \dimen 4 by \dimen 2
				\else	\advance \dimen 4 by -\dimen 2
				\fi
		\repeat
	\fi		
			\xdef \sine {\nodimen 4}%
       }}

% Now the Cosine can be calculated easily by calling \Sine
\def\Cosine#1{\ifx\sine\UnDefined\edef\Savesine{\relax}\else
		             \edef\Savesine{\sine}\fi
	{\dimen0=#1\r@dian\advance\dimen0 by 90\r@dian
	 \Sine{\nodimen 0}
	 \xdef\cosine{\sine}
	 \xdef\sine{\Savesine}}}	      
% end of trig stuff
%%%%%%%%%%%%%%%%%%%%%%%%%%%%%%%%%%%%%%%%%%%%%%%%%%%%%%%%%%%%%%%%%%%%

\def\psdraft{
	\def\@psdraft{0}
	%\ps@typeout{draft level now is \@psdraft \space . }
}
\def\psfull{
	\def\@psdraft{100}
	%\ps@typeout{draft level now is \@psdraft \space . }
}

\psfull

\newif\if@scalefirst
\def\psscalefirst{\@scalefirsttrue}
\def\psrotatefirst{\@scalefirstfalse}
\psrotatefirst

\newif\if@draftbox
\def\psnodraftbox{
	\@draftboxfalse
}
\def\psdraftbox{
	\@draftboxtrue
}
\@draftboxtrue

\newif\if@prologfile
\newif\if@postlogfile
\def\pssilent{
	\@noisyfalse
}
\def\psnoisy{
	\@noisytrue
}
\psnoisy
%%% These are for the option list.
%%% A specification of the form a = b maps to calling \@p@@sa{b}
\newif\if@bbllx
\newif\if@bblly
\newif\if@bburx
\newif\if@bbury
\newif\if@height
\newif\if@width
\newif\if@rheight
\newif\if@rwidth
\newif\if@angle
\newif\if@clip
\newif\if@verbose
\def\@p@@sclip#1{\@cliptrue}

\newif\if@decmpr

%%% GDH 7/26/87 -- changed so that it first looks in the local directory,
%%% then in a specified global directory for the ps file.
%%% RPR 6/25/91 -- changed so that it defaults to user-supplied name if
%%% boundingbox info is specified, assuming graphic will be created by
%%% print time.
%%% TJD 10/19/91 -- added bbfile vs. file distinction, and @decmpr flag

\def\@p@@sfigure#1{\def\@p@sfile{null}\def\@p@sbbfile{null}
	        \openin1=#1.bb
		\ifeof1\closein1
	        	\openin1=\figurepath#1.bb
			\ifeof1\closein1
			        \openin1=#1
				\ifeof1\closein1%
				       \openin1=\figurepath#1
					\ifeof1
					   \ps@typeout{Error, File #1 not found}
						\if@bbllx\if@bblly
				   		\if@bburx\if@bbury
			      				\def\@p@sfile{#1}%
			      				\def\@p@sbbfile{#1}%
							\@decmprfalse
				  	   	\fi\fi\fi\fi
					\else\closein1
				    		\def\@p@sfile{\figurepath#1}%
				    		\def\@p@sbbfile{\figurepath#1}%
						\@decmprfalse
	                       		\fi%
			 	\else\closein1%
					\def\@p@sfile{#1}
					\def\@p@sbbfile{#1}
					\@decmprfalse
			 	\fi
			\else
				\def\@p@sfile{\figurepath#1}
				\def\@p@sbbfile{\figurepath#1.bb}
				\@decmprtrue
			\fi
		\else
			\def\@p@sfile{#1}
			\def\@p@sbbfile{#1.bb}
			\@decmprtrue
		\fi}

\def\@p@@sfile#1{\@p@@sfigure{#1}}

\def\@p@@sbbllx#1{
		%\ps@typeout{bbllx is #1}
		\@bbllxtrue
		\dimen100=#1
		\edef\@p@sbbllx{\number\dimen100}
}
\def\@p@@sbblly#1{
		%\ps@typeout{bblly is #1}
		\@bbllytrue
		\dimen100=#1
		\edef\@p@sbblly{\number\dimen100}
}
\def\@p@@sbburx#1{
		%\ps@typeout{bburx is #1}
		\@bburxtrue
		\dimen100=#1
		\edef\@p@sbburx{\number\dimen100}
}
\def\@p@@sbbury#1{
		%\ps@typeout{bbury is #1}
		\@bburytrue
		\dimen100=#1
		\edef\@p@sbbury{\number\dimen100}
}
\def\@p@@sheight#1{
		\@heighttrue
		\dimen100=#1
   		\edef\@p@sheight{\number\dimen100}
		%\ps@typeout{Height is \@p@sheight}
}
\def\@p@@swidth#1{
		%\ps@typeout{Width is #1}
		\@widthtrue
		\dimen100=#1
		\edef\@p@swidth{\number\dimen100}
}
\def\@p@@srheight#1{
		%\ps@typeout{Reserved height is #1}
		\@rheighttrue
		\dimen100=#1
		\edef\@p@srheight{\number\dimen100}
}
\def\@p@@srwidth#1{
		%\ps@typeout{Reserved width is #1}
		\@rwidthtrue
		\dimen100=#1
		\edef\@p@srwidth{\number\dimen100}
}
\def\@p@@sangle#1{
		%\ps@typeout{Rotation is #1}
		\@angletrue
%		\dimen100=#1
		\edef\@p@sangle{#1} %\number\dimen100}
}
\def\@p@@ssilent#1{ 
		\@verbosefalse
}
\def\@p@@sprolog#1{\@prologfiletrue\def\@prologfileval{#1}}
\def\@p@@spostlog#1{\@postlogfiletrue\def\@postlogfileval{#1}}
\def\@cs@name#1{\csname #1\endcsname}
\def\@setparms#1=#2,{\@cs@name{@p@@s#1}{#2}}
%
% initialize the defaults (size the size of the figure)
%
\def\ps@init@parms{
		\@bbllxfalse \@bbllyfalse
		\@bburxfalse \@bburyfalse
		\@heightfalse \@widthfalse
		\@rheightfalse \@rwidthfalse
		\def\@p@sbbllx{}\def\@p@sbblly{}
		\def\@p@sbburx{}\def\@p@sbbury{}
		\def\@p@sheight{}\def\@p@swidth{}
		\def\@p@srheight{}\def\@p@srwidth{}
		\def\@p@sangle{0}
		\def\@p@sfile{} \def\@p@sbbfile{}
		\def\@p@scost{10}
		\def\@sc{}
		\@prologfilefalse
		\@postlogfilefalse
		\@clipfalse
		\if@noisy
			\@verbosetrue
		\else
			\@verbosefalse
		\fi
}
%
% Go through the options setting things up.
%
\def\parse@ps@parms#1{
	 	\@psdo\@psfiga:=#1\do
		   {\expandafter\@setparms\@psfiga,}}
%
% Compute bb height and width
%
\newif\ifno@bb
\def\bb@missing{
	\if@verbose{
		\ps@typeout{psfig: searching \@p@sbbfile \space  for bounding box}
	}\fi
	\no@bbtrue
	\epsf@getbb{\@p@sbbfile}
        \ifno@bb \else \bb@cull\epsf@llx\epsf@lly\epsf@urx\epsf@ury\fi
}	
\def\bb@cull#1#2#3#4{
	\dimen100=#1 bp\edef\@p@sbbllx{\number\dimen100}
	\dimen100=#2 bp\edef\@p@sbblly{\number\dimen100}
	\dimen100=#3 bp\edef\@p@sbburx{\number\dimen100}
	\dimen100=#4 bp\edef\@p@sbbury{\number\dimen100}
	\no@bbfalse
}
% rotate point (#1,#2) about (0,0).
% The sine and cosine of the angle are already stored in \sine and
% \cosine.  The result is placed in (\p@intvaluex, \p@intvaluey).
\newdimen\p@intvaluex
\newdimen\p@intvaluey
\def\rotate@#1#2{{\dimen0=#1 sp\dimen1=#2 sp
%            	calculate x' = x \cos\theta - y \sin\theta
		  \global\p@intvaluex=\cosine\dimen0
		  \dimen3=\sine\dimen1
		  \global\advance\p@intvaluex by -\dimen3
% 		calculate y' = x \sin\theta + y \cos\theta
		  \global\p@intvaluey=\sine\dimen0
		  \dimen3=\cosine\dimen1
		  \global\advance\p@intvaluey by \dimen3
		  }}
\def\compute@bb{
		\no@bbfalse
		\if@bbllx \else \no@bbtrue \fi
		\if@bblly \else \no@bbtrue \fi
		\if@bburx \else \no@bbtrue \fi
		\if@bbury \else \no@bbtrue \fi
		\ifno@bb \bb@missing \fi
		\ifno@bb \ps@typeout{FATAL ERROR: no bb supplied or found}
			\no-bb-error
		\fi
		%
%\ps@typeout{BB: \@p@sbbllx, \@p@sbblly, \@p@sbburx, \@p@sbbury} 
%
% store height/width of original (unrotated) bounding box
		\count203=\@p@sbburx
		\count204=\@p@sbbury
		\advance\count203 by -\@p@sbbllx
		\advance\count204 by -\@p@sbblly
		\edef\ps@bbw{\number\count203}
		\edef\ps@bbh{\number\count204}
		%\ps@typeout{ psbbh = \ps@bbh, psbbw = \ps@bbw }
		\if@angle 
			\Sine{\@p@sangle}\Cosine{\@p@sangle}
	        	{\dimen100=\maxdimen\xdef\r@p@sbbllx{\number\dimen100}
					    \xdef\r@p@sbblly{\number\dimen100}
			                    \xdef\r@p@sbburx{-\number\dimen100}
					    \xdef\r@p@sbbury{-\number\dimen100}}
%
% Need to rotate all four points and take the X-Y extremes of the new
% points as the new bounding box.
                        \def\minmaxtest{
			   \ifnum\number\p@intvaluex<\r@p@sbbllx
			      \xdef\r@p@sbbllx{\number\p@intvaluex}\fi
			   \ifnum\number\p@intvaluex>\r@p@sbburx
			      \xdef\r@p@sbburx{\number\p@intvaluex}\fi
			   \ifnum\number\p@intvaluey<\r@p@sbblly
			      \xdef\r@p@sbblly{\number\p@intvaluey}\fi
			   \ifnum\number\p@intvaluey>\r@p@sbbury
			      \xdef\r@p@sbbury{\number\p@intvaluey}\fi
			   }
%			lower left
			\rotate@{\@p@sbbllx}{\@p@sbblly}
			\minmaxtest
%			upper left
			\rotate@{\@p@sbbllx}{\@p@sbbury}
			\minmaxtest
%			lower right
			\rotate@{\@p@sbburx}{\@p@sbblly}
			\minmaxtest
%			upper right
			\rotate@{\@p@sbburx}{\@p@sbbury}
			\minmaxtest
			\edef\@p@sbbllx{\r@p@sbbllx}\edef\@p@sbblly{\r@p@sbblly}
			\edef\@p@sbburx{\r@p@sbburx}\edef\@p@sbbury{\r@p@sbbury}
%\ps@typeout{rotated BB: \r@p@sbbllx, \r@p@sbblly, \r@p@sbburx, \r@p@sbbury}
		\fi
		\count203=\@p@sbburx
		\count204=\@p@sbbury
		\advance\count203 by -\@p@sbbllx
		\advance\count204 by -\@p@sbblly
		\edef\@bbw{\number\count203}
		\edef\@bbh{\number\count204}
		%\ps@typeout{ bbh = \@bbh, bbw = \@bbw }
}
%
% \in@hundreds performs #1 * (#2 / #3) correct to the hundreds,
%	then leaves the result in @result
%
\def\in@hundreds#1#2#3{\count240=#2 \count241=#3
		     \count100=\count240	% 100 is first digit #2/#3
		     \divide\count100 by \count241
		     \count101=\count100
		     \multiply\count101 by \count241
		     \advance\count240 by -\count101
		     \multiply\count240 by 10
		     \count101=\count240	%101 is second digit of #2/#3
		     \divide\count101 by \count241
		     \count102=\count101
		     \multiply\count102 by \count241
		     \advance\count240 by -\count102
		     \multiply\count240 by 10
		     \count102=\count240	% 102 is the third digit
		     \divide\count102 by \count241
		     \count200=#1\count205=0
		     \count201=\count200
			\multiply\count201 by \count100
		 	\advance\count205 by \count201
		     \count201=\count200
			\divide\count201 by 10
			\multiply\count201 by \count101
			\advance\count205 by \count201
		     \count201=\count200
			\divide\count201 by 100
			\multiply\count201 by \count102
			\advance\count205 by \count201
		     \edef\@result{\number\count205}
}
\def\compute@wfromh{
		% computing : width = height * (bbw / bbh)
		\in@hundreds{\@p@sheight}{\@bbw}{\@bbh}
		%\ps@typeout{ \@p@sheight * \@bbw / \@bbh, = \@result }
		\edef\@p@swidth{\@result}
		%\ps@typeout{w from h: width is \@p@swidth}
}
\def\compute@hfromw{
		% computing : height = width * (bbh / bbw)
	        \in@hundreds{\@p@swidth}{\@bbh}{\@bbw}
		%\ps@typeout{ \@p@swidth * \@bbh / \@bbw = \@result }
		\edef\@p@sheight{\@result}
		%\ps@typeout{h from w : height is \@p@sheight}
}
\def\compute@handw{
		\if@height 
			\if@width
			\else
				\compute@wfromh
			\fi
		\else 
			\if@width
				\compute@hfromw
			\else
				\edef\@p@sheight{\@bbh}
				\edef\@p@swidth{\@bbw}
			\fi
		\fi
}
\def\compute@resv{
		\if@rheight \else \edef\@p@srheight{\@p@sheight} \fi
		\if@rwidth \else \edef\@p@srwidth{\@p@swidth} \fi
		%\ps@typeout{rheight = \@p@srheight, rwidth = \@p@srwidth}
}
%		
% Compute any missing values
\def\compute@sizes{
	\compute@bb
	\if@scalefirst\if@angle
% at this point the bounding box has been adjsuted correctly for
% rotation.  PSFIG does all of its scaling using \@bbh and \@bbw.  If
% a width= or height= was specified along with \psscalefirst, then the
% width=/height= value needs to be adjusted to match the new (rotated)
% bounding box size (specifed in \@bbw and \@bbh).
%    \ps@bbw       width=
%    -------  =  ---------- 
%    \@bbw       new width=
% so `new width=' = (width= * \@bbw) / \ps@bbw; where \ps@bbw is the
% width of the original (unrotated) bounding box.
	\if@width
	   \in@hundreds{\@p@swidth}{\@bbw}{\ps@bbw}
	   \edef\@p@swidth{\@result}
	\fi
	\if@height
	   \in@hundreds{\@p@sheight}{\@bbh}{\ps@bbh}
	   \edef\@p@sheight{\@result}
	\fi
	\fi\fi
	\compute@handw
	\compute@resv}

%
% \psfig
% usage : \psfig{file=, height=, width=, bbllx=, bblly=, bburx=, bbury=,
%			rheight=, rwidth=, clip=}
%
% "clip=" is a switch and takes no value, but the `=' must be present.
\def\psfig#1{\vbox {
	% do a zero width hard space so that a single
	% \psfig in a centering enviornment will behave nicely
	%{\setbox0=\hbox{\ }\ \hskip-\wd0}
	%
	\ps@init@parms
	\parse@ps@parms{#1}
	\compute@sizes
	\ifnum\@p@scost<\@psdraft{
		\special{ps::[begin] 	\@p@swidth \space \@p@sheight \space
				\@p@sbbllx \space \@p@sbblly \space
				\@p@sbburx \space \@p@sbbury \space
				startTexFig \space }
		\if@angle
			\special {ps:: \@p@sangle \space rotate \space} 
		\fi
		\if@clip{
			\if@verbose{
				\ps@typeout{(clip)}
			}\fi
			\special{ps:: doclip \space }
		}\fi
		\if@prologfile
		    \special{ps: plotfile \@prologfileval \space } \fi
		\if@decmpr{
			\if@verbose{
				\ps@typeout{psfig: including \@p@sfile.Z \space }
			}\fi
			\special{ps: plotfile "`zcat \@p@sfile.Z" \space }
		}\else{
			\if@verbose{
				\ps@typeout{psfig: including \@p@sfile \space }
			}\fi
			\special{ps: plotfile \@p@sfile \space }
		}\fi
		\if@postlogfile
		    \special{ps: plotfile \@postlogfileval \space } \fi
		\special{ps::[end] endTexFig \space }
		% Create the vbox to reserve the space for the figure.
		\vbox to \@p@srheight sp{
		% 1/92 TJD Changed from "true sp" to "sp" for magnification.
			\hbox to \@p@srwidth sp{
				\hss
			}
		\vss
		}
	}\else{
		% draft figure, just reserve the space and print the
		% path name.
		\if@draftbox{		
			% Verbose draft: print file name in box
			\hbox{\frame{\vbox to \@p@srheight sp{
			\vss
			\hbox to \@p@srwidth sp{ \hss \@p@sfile \hss }
			\vss
			}}}
		}\else{
			% Non-verbose draft
			\vbox to \@p@srheight sp{
			\vss
			\hbox to \@p@srwidth sp{\hss}
			\vss
			}
		}\fi

	}\fi
}}
\psfigRestoreAt
\let\@=\LaTeXAtSign

\title{Dynamics of a single particle in a horizontally shaken box}
\author{Barbara Drossel and Thomas Prellberg \footnote{email:
drossel@a13.ph.man.ac.uk, prel@a13.ph.man.ac.uk}} \address{Department
of Theoretical Physics, University of Manchester, Manchester M13 9PL,
UK} \date{\today} \maketitle
\begin{abstract}
We study the dynamics of a particle in a horizontally and periodically
shaken box as a function of the box parameters and the coefficient of
restitution. For certain parameter values, the particle becomes
regularly chattered at one of the walls, thereby loosing all its
kinetic energy relative to that wall. The number of container
oscillations between two chattering events depends in a fractal manner
on the parameters of the system. In contrast to a vertically vibrated
particle, for which chattering is claimed to be the generic fate, the
horizontally shaken particle can become trapped on a periodic
orbit and follow the period--doubling route to chaos when the
coefficient of restitution is changed. We also discuss the case of a
completely elastic particle, and the influence of friction between the
particle and the bottom of the container.
\end{abstract}
\begin{multicols}{2}

\section{Introduction and summary}
\label{intro}

While vertically shaken granular materials have been the object of
intensive research in the past years, the investigation of
horizontally shaken granular materials has just started
\cite{str96,ris97}.  Vertically shaken materials show various cellular
patterns and localized oscillons \cite{mel94,umb96,cle96}, and their
horizontal counterpart was recently found to display ripple--like
patterns \cite{str96}.  Since these patterns are due to the collective
behaviour of many interacting particles, the one--particle system shows
completely different phenomena that are, however, equally fascinating.
The dynamical evolution of a bouncing ball on a vibrating platform was
studied in \cite{meh90,luc93}. As long as the coefficient of
restitution is smaller than one, a particle that hits the platform
with sufficiently small relative velocity bounces off the platform
infinitely often during a finite time and looses its memory of earlier
dynamics. The authors of \cite{luc93} argue that this ``chattering''
is the fate of generic trajectories, which therefore become
periodic. They conclude that true chaos cannot be observed in this
system.

In this paper, we study the dynamics of a singe particle in a
horizontally shaken box. While chattering occurs for part of the
parameter values and initial conditions, we find as well other generic
scenarios like periodic orbits without chatter, the period--doubling
route to chaos, and strange attractors. The interplay between these
different modes of behaviour makes this apparently simple system
astonishingly rich and fascinating.

The outline of this paper is as follows: In the next section we define
the system used in our simulations. In section \ref{inelastic} we
discuss the limiting case of a completely inelastic particle that assumes
the velocity of the wall at each collision. When the particle hits the
wall during the half period where the wall accelerates toward it, it
sticks to the wall until the end of the half period. The number of
reflections until this locking occurs depends in a fractal manner on
the parameter of the system. We also discuss the influence of friction
between the particle and the bottom of the container.  Next, the
opposite limit of a completely elastic particle is considered in
section \ref{elastic}. This system displays all signatures of
Hamiltonian chaotic systems, including periodic, quasiperiodic, and
chaotic orbits.  The physically most relevant case of a partially
elastic particle is then studied in section \ref{partelastic}. On the
one hand, increasing the coefficient of restitution from zero, the
period--doubling route to chaos is observed in many cases. Ultimately,
the strange attractor becomes so large that it includes the chattering
region, thus making the orbits again periodic. On the other hand,
decreasing the coefficient of restitution from one, the neutrally stable
fixed points become attractive. All irrational tori disappear and all 
trajectories appear to first become periodic. A trajectory starting 
in the locking region may or may not lead back 
to it, and several periodic orbits coexist, their basins of attraction 
being strongly interwoven.  Section \ref{conclusion} concludes the paper.

\section{The model}
\label{themodel}
The left and right wall of a horizontally shaken container of length
$L$ are described by the equations
\begin{eqnarray*}
x_{\text{leftwall}}& =& A \sin(\omega t)\\
x_{\text{rightwall}}& =& L +  A \sin(\omega t) \, .
\end{eqnarray*}
$\omega$ is the frequency, and $A$ the amplitude of shaking.  We
denote position and velocity of the particle by $x$ and $v$
respectively, and introduce the relative position and velocity with
respect to the container walls \beq l=x-A\sin(\omega t)\quad,\quad
u=v-A\omega\cos(\omega t)\;.  \eeq Between collisions, the particle
moves according to a linear friction law $dv/dt=-\gamma u$, i.e.,  \beq
du/dt=-\gamma u+A\omega^2\sin(\omega t) \, ,\eeq with a friction
coefficient $\gamma$ which will be set equal to zero in most parts of
this paper.

The collisions with the wall are partially inelastic, and the relative
velocity $u$ changes according to \beq u'=-\eta u \eeq at each
collision, where $\eta$ is the coefficient of restitution.

It is convenient to measure the particle position in units of the
amplitude $A$, time in units of the inverse frequency $\omega^{-1}$,
and velocity in units of $A\omega$. Then the system can be described
by the dimensionless parameters $\alpha=L/A$,
$\tilde\gamma=\gamma/\omega$, and $\eta$. We denote the dimensionless
time, length, and velocity again by $t$, $l$, and $u$. We also
introduce the phase of the container oscillation $\phi=\omega t \hbox{
mod } 2\pi$.

The motion of the particle can be written down immediately. It is most
convenient to describe the particle dynamics in terms of a map that
gives the phase of the container oscillation and the particle velocity
immediately after an impact as function of their values immediately
after the previous impact. Let the particle leave the left wall
($l=0$) after an impact at $t=t_0$ with a relative velocity $u_0$.
The particle moves according to
\begin{eqnarray}\label{movingwithfriction}
l_{t_0,u_0}(t)&=&{1\over\tilde\gamma}(\cos t_0+u_0)
-{1\over\tilde\gamma^2+1}(\sin t+\tilde\gamma\cos t) \nonumber \\
&&-\left(u_0-{1\over\tilde\gamma^2+1}(\tilde\gamma\sin t_0-\cos t_0)\right)
\nonumber \\
&&\times {1\over\tilde\gamma}\exp(-\tilde\gamma(t-t_0))\,,
\end{eqnarray}
the corresponding velocity being $u=dl/dt$. We will mainly discuss $\gamma=0$,
where this equation reduces to 
\beq
l_{t_0,u_0}(t)=\sin t_0-\sin t+
(u_0+\cos t_0)(t-t_0)\;.
\label{nofric}
\eeq
%\beq
%u_{t_0,u_0}(t)={1\over\tilde\gamma^2+1}(\tilde\gamma\sin t-\cos t)+
%\left(u_0-{1\over\tilde\gamma^2+1}(\tilde\gamma\sin t_0-\cos t_0)\right)
%\exp(-\tilde\gamma(t-t_0))
%\eeq

The next collision occurs at the smallest solution $t_c>t_0$ of
\begin{subeqnarray}
l_{t_0,u_0}(t_c)&=&0\qquad\mbox{impact at same wall} \slabel{impsame}\\ 
l_{t_0,u_0}(t_c)&=&\alpha \qquad\mbox{impact at other wall}\slabel{impother}
\end{subeqnarray}
at which the velocity is $u_c=u_{t_0,u_0}(t_c)$.  If the particle
impacts again the same wall, the new velocity is now $u_1=-\eta u_c$.
On the other hand, if the particle collides with the right wall, it is
convenient to use the symmetry between left and right wall to map the
right wall to the left wall via $t\rightarrow t-\pi$ and
$l\rightarrow\alpha-l$, so that now $u_1=\eta u_c$. The velocity is
thus always measured with respect to the wall at which the last impact
has occurred.

Given an initial phase 
$\phi_0=t_0$ and velocity $u_0$, we thus arrive at the map
\begin{subeqnarray}
\phi_1&=&t_c \mod2\pi\, ,\quad u_1=-\eta u_c \phantom{33333}\nonumber\\
&&\qquad\qquad\qquad\mbox{  impact at same wall}\slabel{modsame}\\
\phi_1&=&(t_c-\pi) \mod 2\pi\, ,
\quad u_1=\eta u_c\phantom{33333}\nonumber\\
&&\qquad\qquad\qquad \mbox{  impact at other wall}\slabel{modother}
\end{subeqnarray}

Before we will study this mapping in detail below, we shall briefly discuss
the case of no dissipation ($\eta=1$ and $\gamma=0$). Here,
the dynamics can be derived from a
time--dependent Hamiltonian \cite{lin86,rei92}
\beq
H={p^2\over2m} + V_{SQ}(q) + A m \omega^2 q \sin(\omega t)\, ,
\label{hamiltonian}
\eeq where $ V_{SQ}(q)$ is a square--well potential describing the
walls at $q=0$ and $L$, and $m$ and $p$ are mass and momentum of the
particle. The last term in the Hamiltonian (\ref{hamiltonian}) causes
a periodic oscillation of the particle with respect to the
walls. (Since we describe the system in terms of relative coordinates
and velocities between the particle and the wall, a system with
periodically oscillating walls and a system with a periodic particle
oscillation superimposed to the ballistic motion are equivalent.)
Measuring $p$ in units of $mL\omega$, $H$ and $V$ in units of
$mL^2\omega^2$, $q$ in units of $L$, and $t$ in units of
$\omega^{-1}$, , we arrive at a dimensionless formulation, \beq
\label{ham}
H={p^2\over2}+v_{SQ}(q) + {q\over\alpha}\sin t
\eeq
where $v_{SQ}(q)$ is an appropriately scaled square well potential with
walls at $0$ and $1$. Upon introducing action--angle variables 
$(J,\theta)$ for the ``unperturbed'' Hamiltonian $H_0={p^2/2}+v_{SQ}(q)$ 
\cite{lin86,rei92}, the Hamiltonian (\ref{ham}) transforms to
\beq
\label{actionangle}
H={\pi^2J^2\over2}+\alpha^{-1}\left({1\over2}\sin t
+{2\over\pi^2}\sum_{n=-\infty\above0cm n\;odd}^{\infty}
{1\over n^2}\sin(n\theta-t)\right)
\eeq
The transformation between $(p,q)$ and $(J,\theta)$ is given by
\beq
\label{trafo}
p={\pi J}\sign\theta\quad\mbox{and}\quad 
q=|\theta|/\pi 
\eeq
for $-\pi<\theta<\pi$, and periodically continued in $\theta$.
We have included the term  $(2\alpha)^{-1}\sin t$ in the Hamiltonian 
(\ref{actionangle}) for completeness sake, but we shall drop it
below as it does not influence the dynamics.

\section{The completely inelastic particle}
\label{inelastic}
\subsection{Modelling by a one--dimensional map}
\label{model}

We now discuss the case $\gamma=0$ and $\eta=0$ in detail. The
particle moves freely between the walls, and after an impact the relative
velocity $u_0$ with respect to the wall is zero. This means that the 
two--dimensional mapping defined above is in fact reduced to a 
one-dimensional mapping. The subsequent fate of the particle depends 
only on the phase $\phi_0$ at the moment of impact,
as indicated in Fig.~\ref{fate}. 
\begin{figure}  \narrowtext
\vskip -1cm
\centerline{\psfig{file=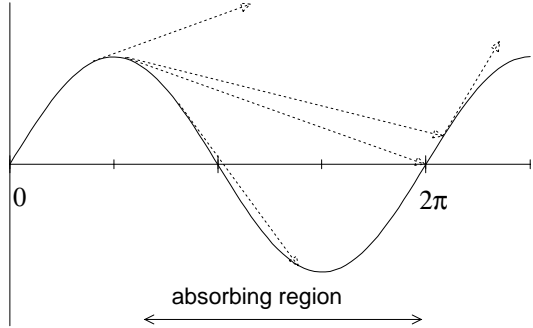,height=2.5in,angle=-90}}
%\vskip 1cm
\caption{The fate of a particle that is inelastically reflected at the wall.}
\label{fate}
\end{figure}

For $\phi_0 \in [0, \pi/2[\mod2\pi$, the particle is reflected from
the wall with (absolute) velocity $\cos(\phi_0)$ and is headed for the
other wall.  For $\phi_0 \in [\pi,2\pi[ \mod 2\pi$, the particle
sticks to the wall until $\phi_0=2\pi$, and then leaves the wall with
(absolute) velocity $1$. In the intermediate region $\phi_0 \in
]\pi/2,\pi[\mod2\pi$, the sign of the particle velocity is not
reversed during the collision, and the particle hits the same wall
again at a later phase $\phi_1$ which is determined by the equations
(\ref{nofric}), (\ref{impsame}), and (\ref{modsame}), leading to
\beq\label{samewall} \sin\phi_0-\sin\phi_1+(\phi_1-\phi_0)\cos\phi_0=0\, .
\eeq 
For $\phi_0 \ge \phi^c = 1.79..$, the phase $\phi_1$ is in the
locking interval $ [\pi,2\pi[$. As we are interested in the dynamics
between impacts at opposite walls, we can extend the locking interval
to include $\phi^c$, i.e., to the interval $[\phi_c,2\pi]$. For
$\phi_0 < \phi^c$, the phase $\phi_1$ is in the interval $[0, \pi/2[
\mod2\pi$ and the particle is immediately reflected towards the other
wall, where it arrives at a phase $\phi_2$ now given by
(\ref{impother}) and (\ref{modother}), i.e., \beq\label{otherwall}
\sin\phi_1+\sin\phi_2+(\phi_2-\pi-\phi_1)\cos\phi_1=\alpha\, . \eeq 

From now on, let $\phi_i$ denote the phase of a wall at the moment
where the particle is reflected towards the other wall for the $i$th
time. We do not count reflections that lead to a subsequent reflection
at the same wall. Then, the map that gives $\phi_{i+1}$ as function of
$\phi_i$ is a map of the interval $[0,\pi/2[$ into itself, dependent
only on the dimensionless parameter $\alpha=L/A$. The map
$\phi_{i+1}=\Phi_\alpha(\phi_i)$ is given implicitly by the equations
(\ref{samewall}) and (\ref{otherwall}), and is shown in Fig.~\ref{map}
for the value $\alpha=5$. Branches with positive slope correspond to a
particle arriving at the other wall at a phase in the interval
$[0,\pi/2]$, from where it is immediately reflected back, while
branches with negative slope indicate a twofold reflection at the
other wall, the first reflection being in the phase interval
$]\pi/2,\phi^c]$.  The dashed line is the diagonal
$\phi_{i+1}=\phi_i$, and the dotted line indicates the fate of a
particle starting at phase zero.  For $\pi/2>\phi_i>\pi-\phi^c$, the
velocity of the particle is so small that it cannot hit the other wall
in the locking region. A small interval $\Delta\phi_i$ maps onto an
interval $\Delta \phi_{i+1}$ proportional to the flight time, which is
of the order $\alpha / \cos\phi_i$. For this reason, the density of
branches in the map increases with increasing $\alpha$ and $\phi_i$
and diverges when $\phi_i$ approaches the value $\pi/2$, where the
initial velocity decreases to zero.
\begin{figure}  \narrowtext
\centerline{\psfig{file=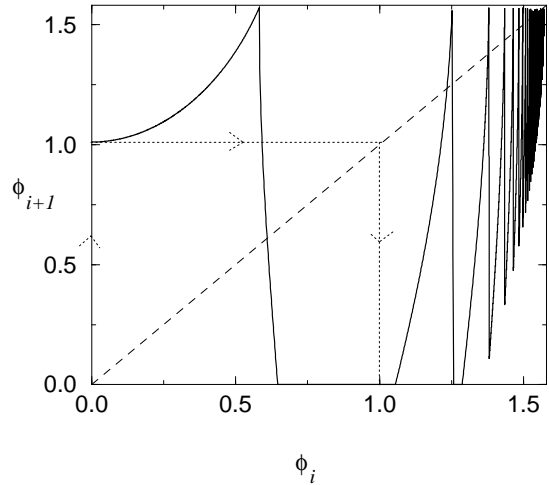,height=2.5in,angle=-90}}
%\vskip 1cm
\caption{The map for $\alpha=5$.}
\label{map}
\end{figure}

\subsection{Periodic trajectories}
\label{periodictrajectiories}

The phase $\Phi_\alpha(0)$ at which a particle starting at phase zero 
will be reflected from the other wall depends on $\alpha$. It is zero for
$\alpha \in [(2n-1)\pi+\phi^c+\sin\phi^c, (2n+1)\pi]$, for any
nonnegative integer $n$. In this case, a particle that starts in the
locking region will hit the locking region at the other wall, and the
periodicity of such a trajectory is one. For all values of $\alpha$,
the map intersects the perpendicular axis with a slope zero, since the
initial velocity does not change with $\phi_i$ for $\phi_i=0$. For
values $\alpha$ only slightly above $(2n+1)\pi$, the map therefore has
a stable fixed point $\phi^\star$ close to zero. This fixed point
vanishes with increasing $\alpha$ via a saddle--node bifurcation, and
for values of $\alpha$ slightly beyond the bifurcation, a particle
trajectory can be trapped for a long time in the neighborhood of the
former fixed point, before it escapes and hits ultimately the locking
region where $\phi_{i+1}=0$. Due to these locking regions, there exist
no truly chaotic trajectories, but each trajectory has a finite
periodicity.  When $\alpha$ increases further, the number of
reflections in the neighborhood of the former fixed point (or, more
precisely, on the first branch that has positive slope), decreases in
steps of size one. Close to such a decrease, a trajectory that leaves
the first branch goes through the upper right--hand corner of the map
and can therefore have an arbitrarily large periodicity. Further away from 
it, a trajectory can hit the locking region immediately after leaving the
first branch.  In general, very complex trajectories can occur.
%that are not worthwhile to be analyzed in detail. 
In particular when $\Phi_\alpha(0)$ is close to $\pi/2$, the trajectory 
will spend some time in the upper right--hand corner of the map, and 
a slight change in $\alpha$ may lead to a large change in the 
periodicity of the trajectory.

There are also many possibilities for obtaining low--periodicity
trajectories. In Fig.~\ref{map}, e.g., the periodicity is two. A
period--doubling bifurcation sequence occurs when $\alpha$ is
decreased from $(2n-1)\pi+\phi^c+\sin\phi^c$.  In
Fig.~\ref{bifurcations}, this scenario is sketched. 
Since it involves
only the first two branches of the map, the other branches are not
shown. As long as the first branch is small (e.g., for
$\alpha=5.8$), the particle hits the locking region after the second
reflection.  With decreasing $\alpha$, the endpoint of the trajectory
moves to the right, and finally hits the foot of the second branch,
leading to a period--doubling bifurcation. 
For $\alpha=5.743$, one
observes therefore a cycle of period 4. When $\alpha$ is decreased
further, the end point of the 4--cycle moves to the left and hits the
foot of the first branch, leading to another period doubling.  
\begin{figure}  \narrowtext
\centerline{\psfig{file=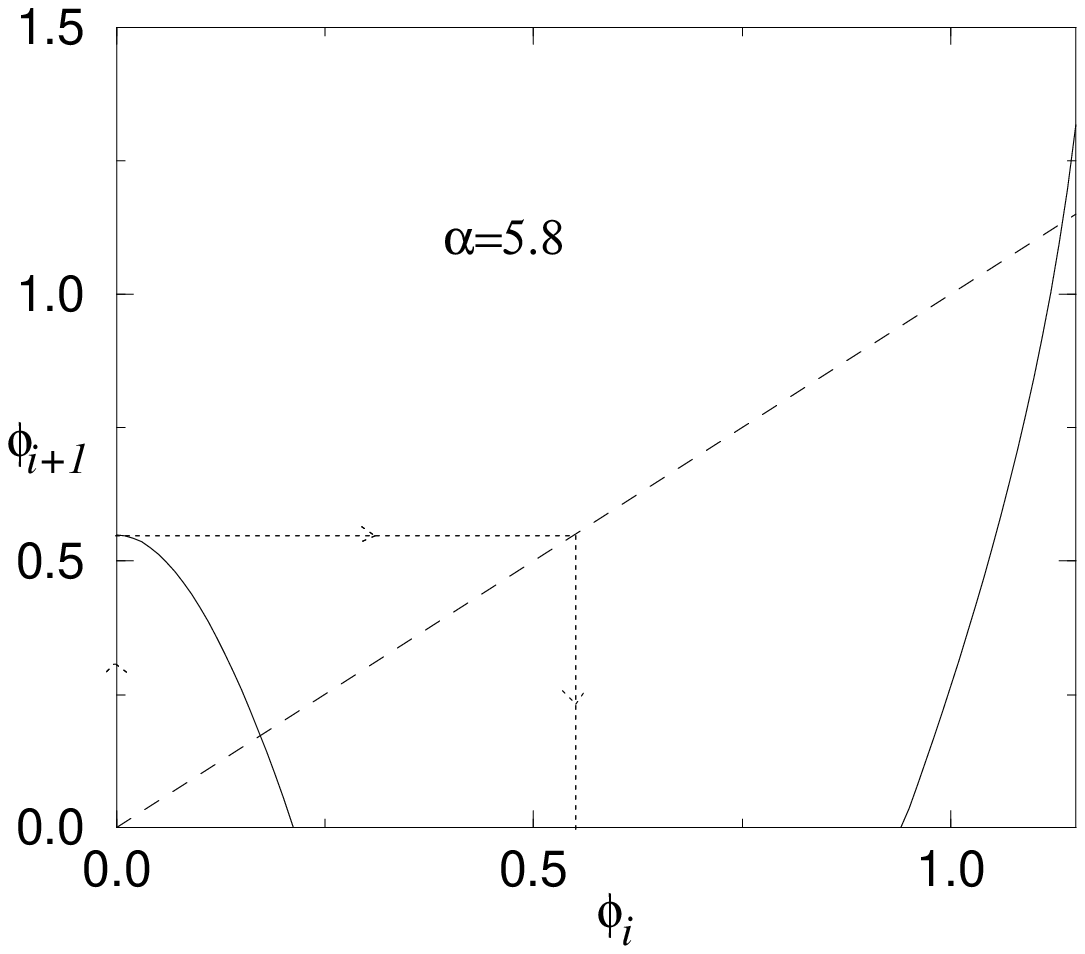,height=2.5in,angle=-90}}
\vskip 0.2in
\centerline{\psfig{file=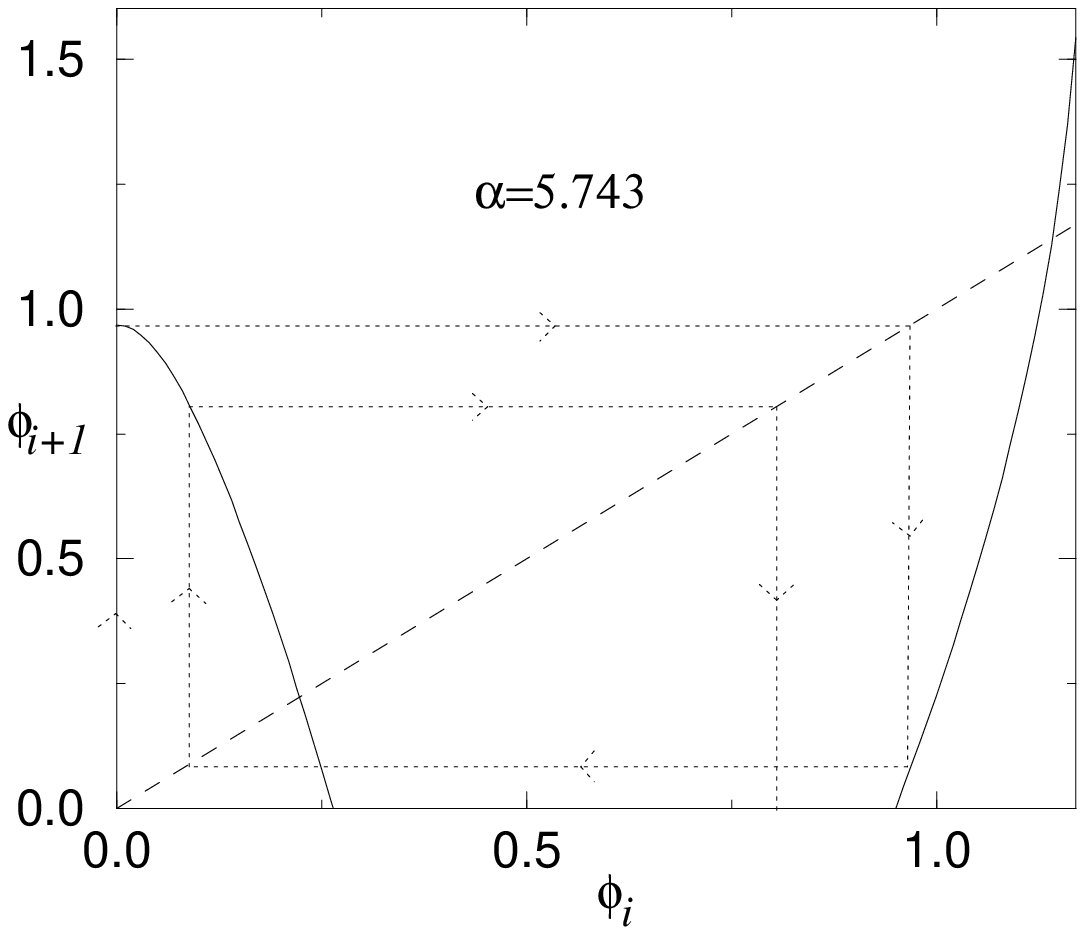,height=2.5in,angle=-90}}
\vskip 0.2in
\centerline{\psfig{file=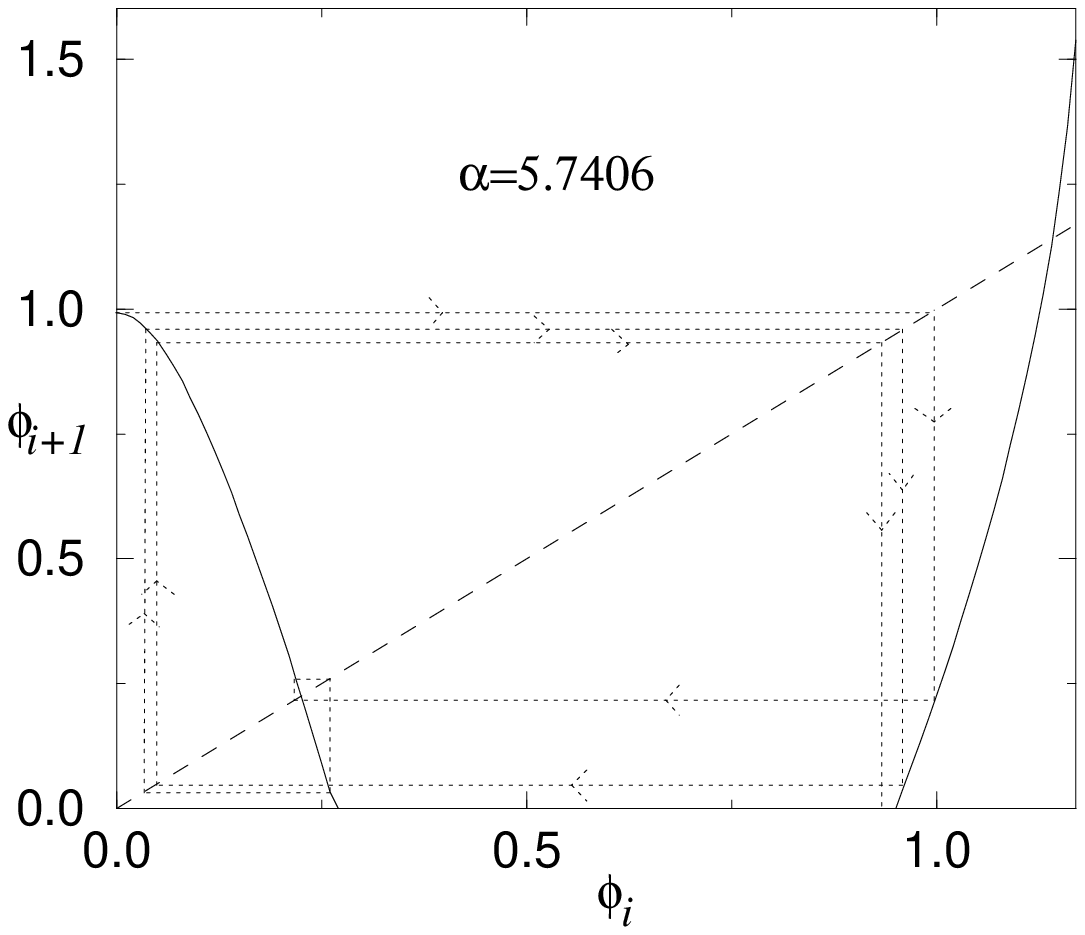,height=2.5in,angle=-90}}
\caption{A series of bifurcations when $\alpha$ is decreased.}
\label{bifurcations}
\end{figure}
The
last part of Fig.~\ref{bifurcations} shows an 8--cycle for
$\alpha=5.7406$. This period--doubling scenario continues as the end
point of the 8--cycle moves to the left, etc., and the period becomes
infinite when the trajectory hits the unstable fixed point on the
first branch. For even smaller values of $\alpha$, other
periodicities are observed that are not powers of 2.

Fig.~\ref{period} shows the period of a trajectory originally starting
in the absorbing region as function of $\alpha$. 
\begin{figure}  \narrowtext
\centerline{\psfig{file=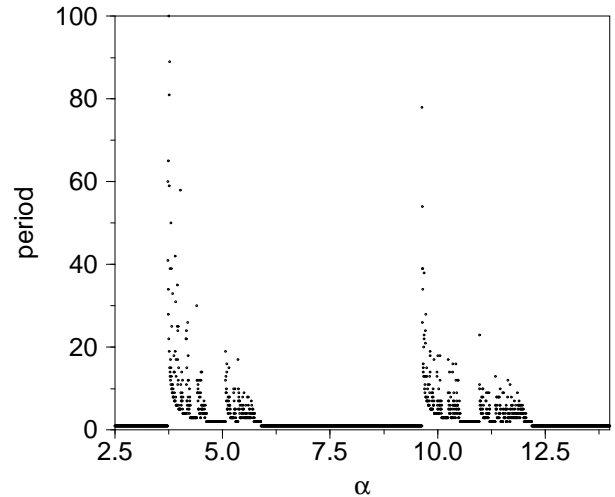,height=2.7in,angle=-90}}
%\vskip 1cm
\caption{Period of trajectories originally starting in the absorbing 
region as function of $\alpha$.}
\label{period}
\end{figure}
All the
above--mentioned features can be seen: The large plateaus of period
1 that extend beyond $(2n+1)\times \pi$ because of the existence of
stable periodic orbits outside the locking region; the stepwise
decrease of the number of subsequent reflections at the first branch
as $\alpha$ increases, and a high periodicity in between these
subsequent plateaus; the trajectory of period 2 shown in
Fig.~\ref{map}; the period--doubling sequence at the left end of the
period--1 plateaus (due to the finite resolution in $\alpha$ only the
2-- and 4--cycles can be seen. The 3--cycle lies beyond the
period--doubling sequence).  As the branches of the map become steeper
and more numerous with increasing $\alpha$, plateaus will become
shorter, and the structures will become more complex for large $\alpha$.

\subsection{Comparison to a vertically shaken particle}
\label{vertical}

In \cite{meh90,luc93}, the periodicity of trajectories of vertically
shaken completely inelastic particles is studied. There, gravitation
brings the particle back to the vibrating platform, and no second wall
is needed. Therefore, fast particles have longer flight times between
reflections than slow particles, while the opposite holds for
horizontally shaken particles.  For large oscillation frequencies, the
authors of \cite{meh90} obtain an approximate map of the form
$\phi_{i+1} \simeq \phi_i + const.\times \cos\phi_i$.  An approximate
map for our system is obtained from Eq.~(\ref{otherwall}) for large
$\alpha$ and reads $\phi_{i+1} \simeq
\phi_i+\pi+\alpha/\cos\phi_i$. The map in \cite{meh90} has no stable
fixed points and no region where the trajectory can be temporarily
trapped. Nevertheless, there are infinite hierarchies of periodic
trajectories between low--period plateaus, leading to a structure on
all scales of the period--vs--frequency plot, as for the horizontally
shaken particle.

\subsection{The influence of friction}
\label{friction}

In the presence of friction between the particle and the container
bottom, there exist stable low--velocity particle trajectories that
never touch the walls, but move periodically back and forth in a small
region somewhere in the central part of the container.  Particles that
are reflected at a phase near $\pi/2$ have a very small velocity and
can become trapped in such a region. Apart from the locking sections
in the map Fig.~\ref{map}, there is consequently an absorbing section
for $\phi_i$ above some threshold that depends on $\alpha$ and the
strength of friction. From equation (\ref{movingwithfriction}) with
$u_0=0$ we see that the maximum possible value of $l$ is \beq
l_{\phi_0,0}^{\text{max}}={1\over\tilde\gamma}\cos\phi_0
+{1\over\sqrt{1+\tilde\gamma^2}}\;.  \eeq The first term is the
initial velocity divided by the friction coefficient, and is the
distance that a particle with initial velocity $\cos(\phi_0)$ can
travel on a stationary surface. The second term is the amplitude of
the periodic oscillation that a particle performs when it is
periodically driven.  If $l_{\phi_0,0}^{\text{max}}<\alpha$,  the
particle never reaches the other wall. If $\alpha>1$, this is
always the case for $\phi_0$ close enough to $\pi/2$, as the
initial (absolute) velocity of the particle is arbitrarily small. If
$\alpha$ is large enough, i.e., \beq
\alpha>{1\over\tilde\gamma}+{1\over\sqrt{1+\tilde\gamma^2}}\;, \eeq
then even a particle that was reflected at phase zero, i.e., with
maximum velocity, does not reach the other wall.
Fig.~\ref{periodfric} shows the period of trajectories as function of
$\alpha$ for $\tilde\gamma=0.025$. 
\begin{figure}  \narrowtext
\centerline{\psfig{file=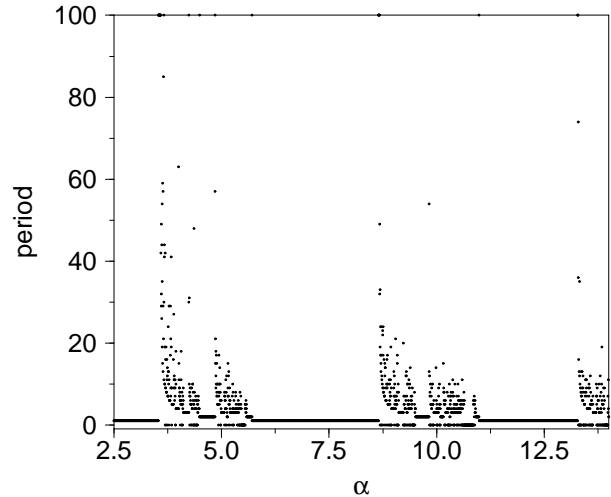,height=2.7in,angle=-90}}
%\vskip 1cm
\caption{Period of trajectories originally starting in the absorbing 
region as function of $\alpha$, with a
friction coefficient $\tilde\gamma=0.025$.}
\label{periodfric}
\end{figure}
Trajectories that get stuck in the
absorbing region are assigned a period zero. To keep the computer
program simple, friction was only active for $x$--coordinates between
$1$ and $\alpha-1$, and only leading terms in $\tilde\gamma$ were
considered.  Comparing Fig.~\ref{periodfric} with Fig.~\ref{period},
one finds that with increasing $\alpha$ the graph becomes more and
more compressed horizontally, and that an increasing fraction of
trajectories get stuck. The region $\alpha>41.0$ where all
trajectories get stuck, is not shown on the graph.

\section{The completely elastic particle}
\label{elastic}

We now study the non-dissipative case $\eta=1$ and $\gamma=0$. Here, the
particle moves without friction and is reflected elastically at the 
container walls, where its relative velocity with respect to the wall 
changes sign at each collision.
The velocity is now an independent variable, adding a new
dimension to the phase space. As the system is non-dissipative,
it can be described by a time--dependent Hamiltonian as described in section
\ref{themodel}. An equivalent problem is studied in
\cite{lin86,rei92,fuk95}, namely the motion of a charged particle in a
one--dimensional infinite square well potential driven by an
oscillating external field. In those papers, the particle performs
periodic oscillations (superimposed to the free motion). The parameter 
$\alpha^{-1}$ is proportional to the electric field. In terms of
coordinates and velocities relative to the container walls, the
dynamics is the same as for our system. 

Since the system
is hamiltonian, phase space volume is conserved in time, and a
stroboscopic Poincare section that gives $l(t)$ and $u(t)$ in time intervals
$T=2\pi$ is area preserving.  However, we find it physically
more relevant to plot the phase of the oscillation and the velocity of
the particle relative to the wall immediately after each
reflection, i.e. $\phi=t_c\mod2\pi$ and $u(t_c)$ for $l(t_c)=0$. 
As the time intervals are not constant, this representation is not 
area preserving, but describes of course the same physics as a true 
Poincare section.
Fig.~\ref{poincare} shows our data points for $\alpha=1.65$,
$\alpha=16.5$, and $\alpha=165$, and for several different
trajectories. The value of $\alpha$ used in \cite{lin86,rei92,fuk95}
is of the order of several hundred, since the authors were mainly
interested in the limit of small electrical field.  

In order to understand some important features of Fig.~\ref{poincare}, 
we turn to the action-angle variable representation (\ref{actionangle}),
i.e.
$$
H={\pi^2J^2\over2}+{2\over\pi^2\alpha}
\sum_{n=-\infty\above0cm n\;odd}^{\infty}{1\over n^2}\sin(n\theta-t)\;.
$$

If one now attempts a canonical
transformation such that the new action and the new Hamiltonian are
constants of motion to first order in $\alpha^{-1}$, one finds that
this perturbation theory breaks down near the {\it resonances}, where
$\pi^2 J n = 1$, i.e., $u=\alpha/\pi n$, for any positive $n$. For
large $\alpha$, the resonances to lowest $n$ (i.e., to highest
energy), are widely separated, and in the neighborhood of such a
resonance, the system is well approximated by a single resonance
Hamiltonian
\beq
H_n ={\pi^2J^2\over2} + {2\over\pi^2\alpha}{\sin(n\theta-t)\over n^2}\; .
\eeq
This Hamiltonian is integrable and becomes identical to the Hamiltonian for
a pendulum under a canonical transformation that equals $\theta-t$
with the pendulum angle variable. For $\alpha=165$,
Fig.~\ref{poincare} shows clearly these pendulum--like
trajectories. In \cite{lin86}, it was shown that the breakdown of the
irrational tori between two resonance regions and the onset of
large--scale chaos can be approximately determined from a
two--resonance Hamiltonian that includes the two resonances adjacent
to the considered torus.  This chaotic region can be seen in the lower
part of all three Figs.~\ref{poincare}.  
The stable periodic orbits
that can be seen for the larger two values of $\alpha$ represent
trajectories that hit the walls always at the moment of farthest
extension, as illustrated in Fig.~\ref{periodicorbit}, and they are at
the center of resonance zones and correspond (in the single resonance
approximation) to a pendulum at stable rest.  
The velocity of such a
trajectory is given by 
\beq
\label{fixedpoint}
u_n^s={\alpha+2\over(2n-1)\pi}\;.
\eeq
The stability of these trajectories is found from linear stability analysis.
The eigenvalues of the stability matrix are
\beq
\lambda_n=1-{\pi\over u_n^s}\pm\sqrt{{\pi\over u_n^s}({\pi\over u_n^s}-2)}\; .
\label{eigenvalues}
\eeq
\begin{figure}  \narrowtext
\centerline{\psfig{file=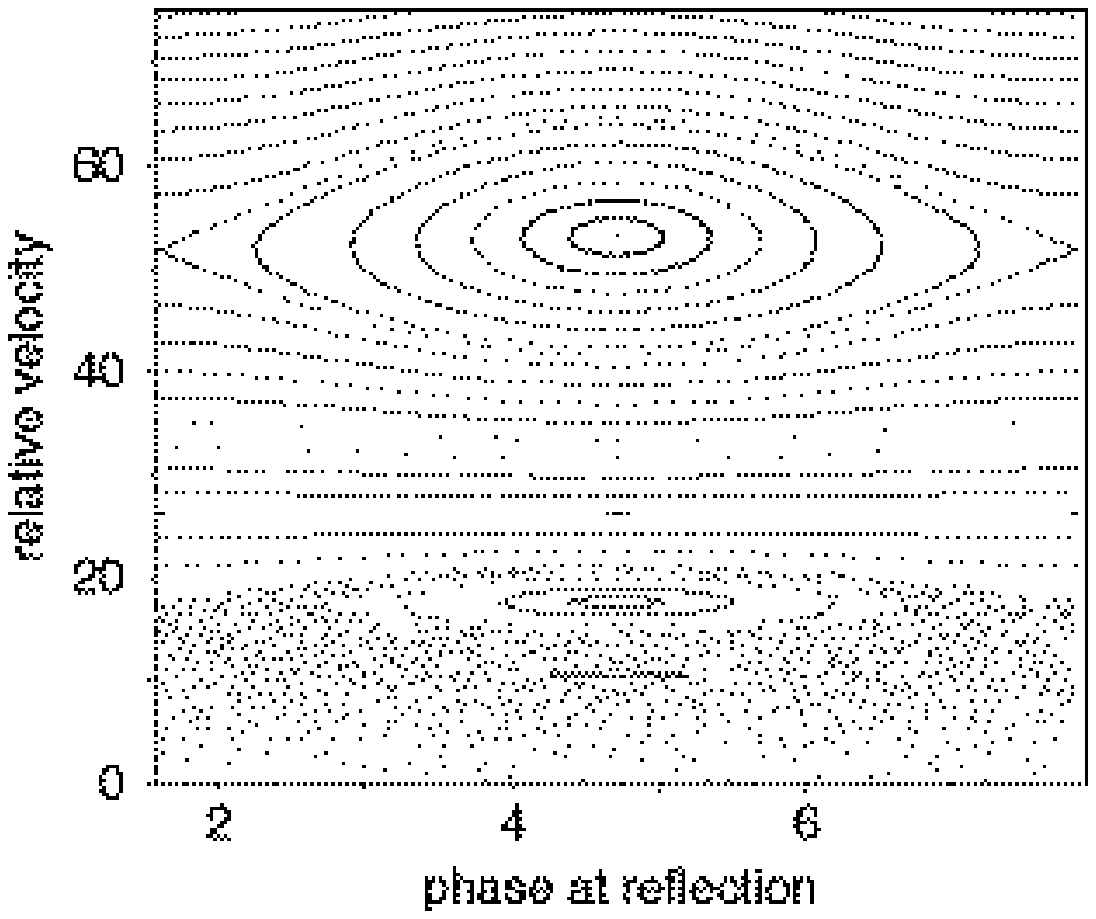,height=2.5in,angle=-90}}
\centerline{\psfig{file=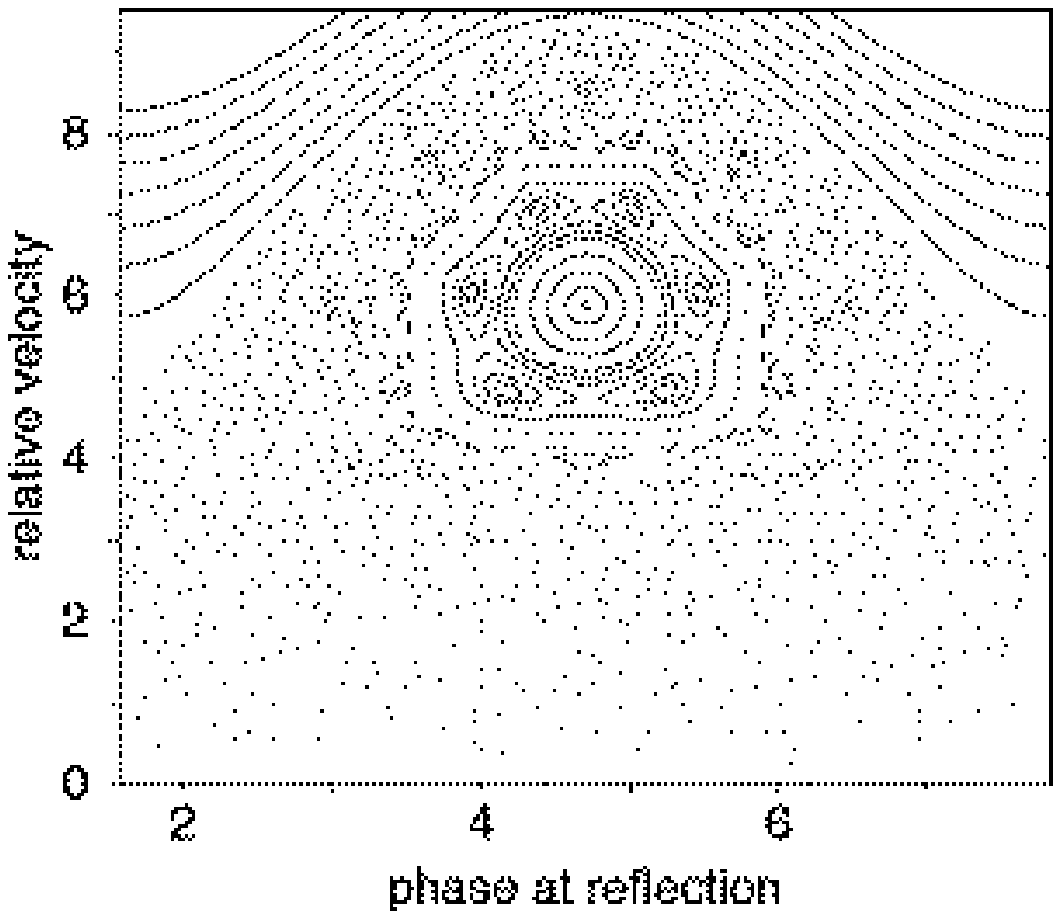,height=2.5in,angle=-90}}
\centerline{\psfig{file=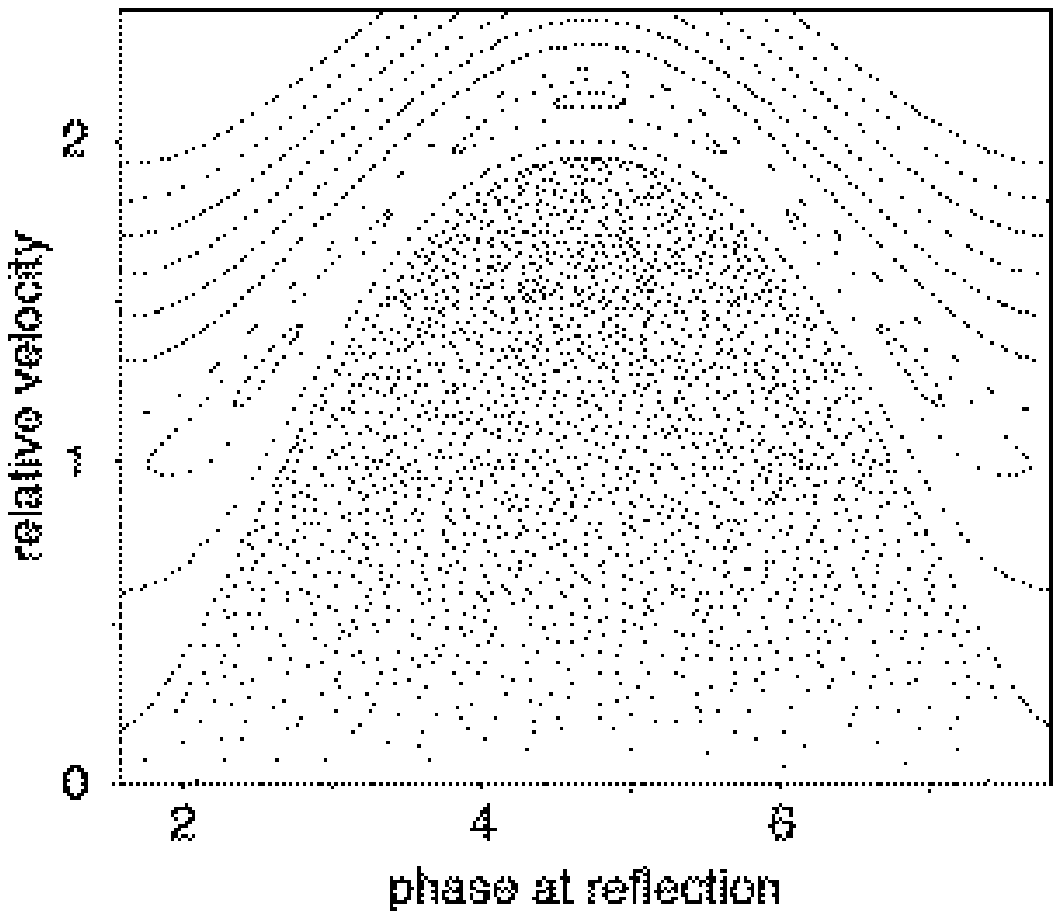,height=2.5in,angle=-90}}
%\vskip 1cm
\caption{Phase and velocity relative to the wall at the moment of
reflection for $\alpha=165$, 16.5, and 1.65, for several different
trajectories. To each trajectory, several 100 points are plotted.}
\label{poincare}
\end{figure}
\begin{figure}  \narrowtext
\centerline{\psfig{file=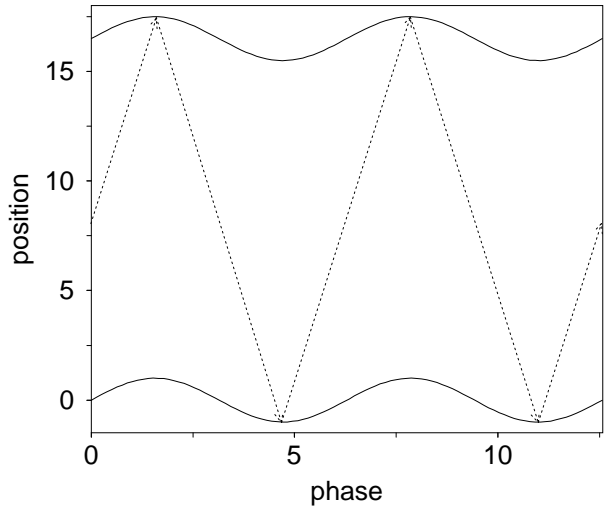,height=2.7in,angle=-90}}
%\vskip 1cm
\caption{The periodic orbit of lowest period.}
\label{periodicorbit}
\end{figure}
For $u_n^s > \pi/ 2$ the eigenvalues are complex conjugate with
$|\lambda|=1$, and the trajectory is neutrally stable. For $u_n^s <
\pi/2$, one eigenvalue is in modulus larger than one, and the trajectory is
unstable.  To each of these stable periodic orbits there exists an unstable
counterpart that hits the walls always at phase $\pi/2$ and that has
the velocity $u_n^u=(\alpha-2)/(2n-1)\pi$. These trajectories
correspond (in the single resonance approximation) to a pendulum at
its unstable fixed point. For nonintegrable systems, unstable periodic
orbits are always part of chaotic regions. It is, however, remarkable,
that the boundary of the large chaotic regions is close to such an
unstable fixed point, and that its shape is close to the shape of a
pendulum separatrix $u \simeq C_1 + C_2\sqrt{1-\sin\phi}$. This means
that the separatrix belonging to the highest--energy resonance within
the chaotic region determines the approximate boundary of the chaotic
region.

For very large velocities $u \gg \alpha$, the phase and velocity
relative to the wall change only by a small fraction between
subsequent reflections, and solving a differential equation to leading
order in $1/u$, one finds on approximate equation for the tori
$$ u \simeq \sqrt{C-\alpha\sin\phi}\, .$$ (More generally, for a
container oscillating according to an equation $x_{\text{wall}} =
f(\phi)$, one finds $ u \simeq \sqrt{C-\alpha f''(\phi)}$.) This 
approximation can also be found from the Hamiltonian (\ref{actionangle}).
A more detailed calculation including the first resonance gives
\beq
u\simeq
\sqrt{\left(\alpha\over\phi\right)^2\left(1+C'\pm
\sqrt{C'-{4\over\alpha}\sin\phi}\right)-\alpha\sin\phi}
\eeq
 
The shape of stable irrational tori for smaller values of the velocity
is obtained in \cite{fuk95} through approximation by periodic
trajectories.

\section{The partially inelastic particle}
\label{partelastic}
The dynamics of the partially elastic particle ($0<\eta<1$) show a
very intricate structure. Depending on the parameters and the initial
state, the trajectory of a particle can be on a low--periodicity
periodic orbit, a strange attractor, or an orbit that looses
periodically all its energy due to ``chattering'', i.e. an infinite
number of reflections at the same wall during a finite time interval.
In the following subsections we discuss all these phenomena and the
interplay between them as the parameter values are varied.

\subsection{Chattering} \label{chatter}
When the particle hits the wall with sufficiently small relative
velocity, it is reflected infinitely often from that wall during a
finite time interval and looses all its energy, as described in
\cite{luc93} for the vertically shaken particle. It subsequently
sticks to the wall until the phase is $\phi=0$, and then leaves the
wall with relative velocity $u=0$. In order to understand this
phenomenon, let us consider first a particle colliding with a wall
that is accelerated at a constant rate $a$. This situation is
equivalent to a ball bouncing in a gravitational field, where it
experiences a constant acceleration toward the ground. If the initial
relative velocity of the particle is $u_0$, the relative velocity
becomes zero after the time
$$T=2\eta u_0/a + 2\eta^2 u_0/a+....=2u_0\eta/a(1-\eta)$$ after the
first collision. In our system, the wall acceleration is of the order
$A\omega^2$, and chattering can occur if the time $T$ is no larger
than of the order of the oscillation period $\omega^{-1}$, leading to
the condition $u_0 \lesssim A\omega (1-\eta)/\eta$ for chattering. For
$\eta$ close to 1, $u_0$ must be very small, which is only possible if
the particle hits the wall at a phase $\phi_0$ within a distance
$\Delta \phi_0 \propto \sqrt{1-\eta}$ of $\pi$, or,
equivalently, if the particle's trajectory hits the (absolute)
position $x=0$ (or $x=L$) within a time interval of the order $\Delta
t \sim u_0\Delta \phi_0 \propto (1-\eta)^{3/2}$. The phase space
volume $u_0 \Delta t$ for which chattering occurs shrinks consequently
as $(1-\eta)^{5/2}$, when $\eta$ approaches 1.

In \cite{luc93}, it is argued that chattering is the generic fate of a
particle bouncing on a vibrating platform, with the number of
reflections between two chattering events diverging as $\eta$
approaches 1. In our system, however, a particle that leaves the wall
at phase $\phi=0$ and with relative velocity $u=0$, returns to the
chattering region only for certain combinations of $\alpha$ and
$\eta$, and is otherwise trapped on a periodic or chaotic orbit. Even
in those cases where a chattered particle is chattered again,
there may exist other trajectories that never enter the chattering
region.

\subsection{The almost elastic case} \label{almost}

Due to dissipation, a given phase space volume shrinks with time. In
particular, regions with large initial velocity are depleted, since
the energy of a particle decreases exponentially fast until it reaches
the regime where the time between impacts is of order unity,
i.e. where its velocity is of order $\alpha$. The change of $u$ with
the impact number $n$ is approximately given by
$$ \hbox{d} u/\hbox{d}n = -u(1-\eta)\,.$$ Using the relation $\hbox{d}
\phi \simeq (\alpha/u)\hbox{d}n$, we find \beq
u\simeq{\alpha\over(1-\eta)(C+\phi)} \eeq for the decrease of $u$ with
$\phi$, with a constant $C$ that is determined by the initial
conditions.  A small initial velocity interval consequently increases
rapidly with increasing $\phi$, leading to a strong shear of an
initial domain of phase space, and interweaving different domains of
attraction, as we shall see below. In fact, every domain of
attraction contains points with arbitrarily large velocity.

Another consequence of the shrinking phase space volume is that all
stable periodic orbits of a Hamiltonian system become attractive when
small dissipation is added \cite{lie85}. This is because the
eigenvalues of the stability matrix remain complex conjugate for
sufficiently small $(1-\eta)$, while their product is $\lambda_1
\lambda_2 = \eta^2$, leading to $|\lambda_i|=\eta < 1$. Thus, for
$\eta$ infinitesimally smaller than $1$, there is a plethora of
attracting periodic orbits with competing domains of attraction. When
$\eta$ decreases further, most of these periodic orbits merge with
unstable orbits and disappear, leaving only few attractive fixed
points. The fixed points in the center of the resonances survive
over a fairly large range of $\eta$ values, as shown by the  
subsequent calculation.  Writing \beq
\rho=(2n-1)\pi{1+\eta\over1-\eta}\;, \eeq the fixed point in the
center of the $n$th resonance changes with decreasing $\eta$ according
to \beq
\label{trajectory}
\sin\phi={2\alpha-\rho\sqrt{4+\rho^2-\alpha^2}\over4+\rho^2}
\quad\mbox{and}\quad u={2\eta\over1-\eta}\cos\phi\;.  \eeq This fixed
point can vanish by merging either with an unstable fixed point, or it
can become unstable.  The merging of stable and unstable fixed points
occurs when the square root in (\ref{trajectory}) becomes imaginary,
leading to the condition \beq
\label{merging}
\eta>{\sqrt{\alpha^2-4}-(2n-1)\pi\over\sqrt{\alpha^2-4}+(2n-1)\pi}
\eeq
for the existence of the fixed point.
\begin{figure}  \narrowtext
\centerline{\psfig{file=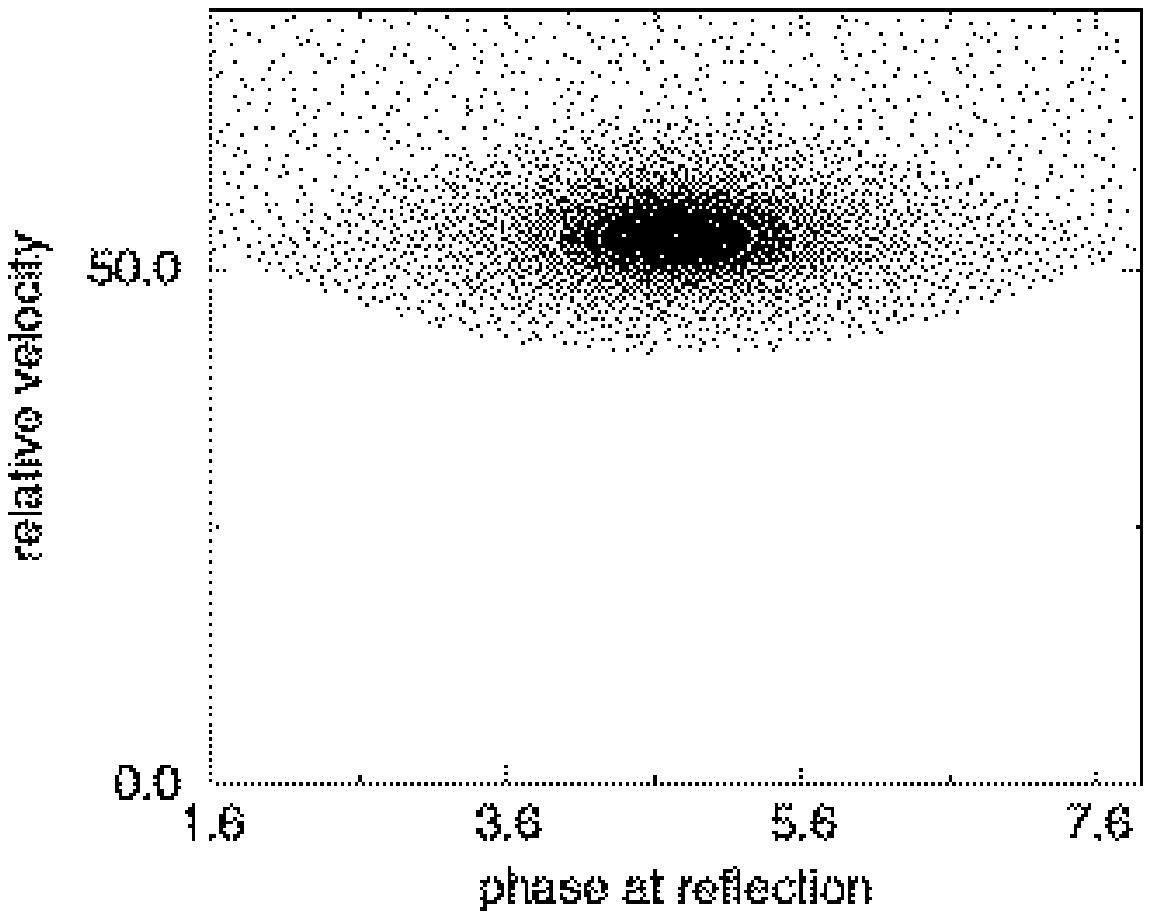,height=2.5in,angle=-90}}
\centerline{\psfig{file=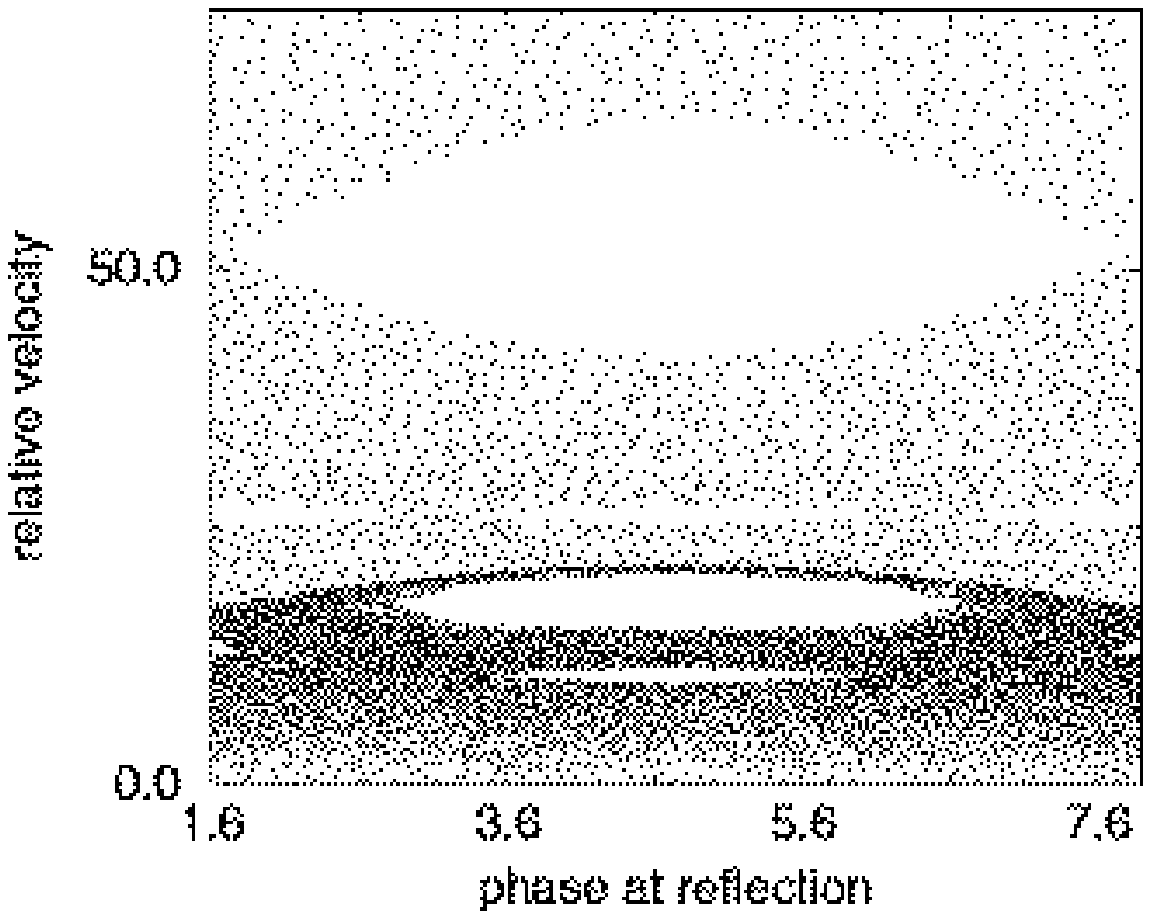,height=2.5in,angle=-90}}
\centerline{\psfig{file=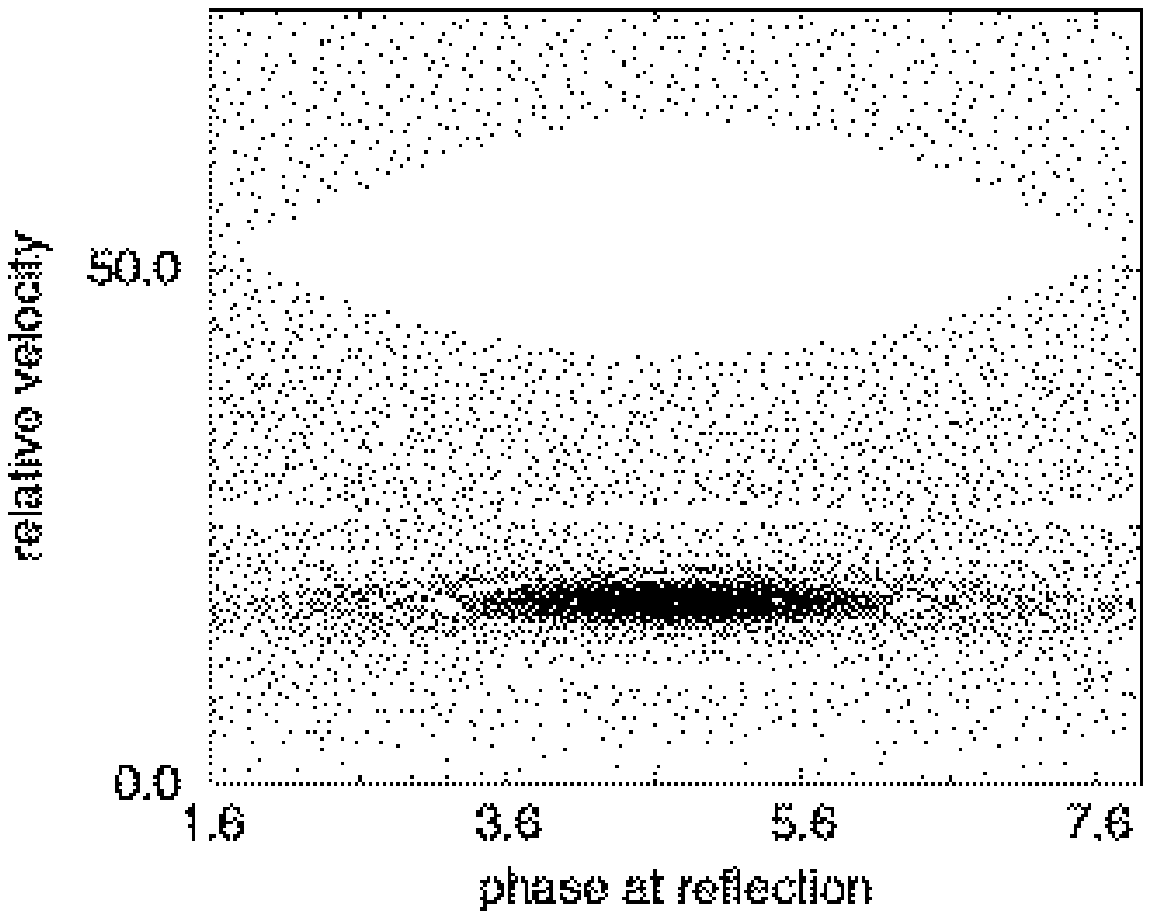,height=2.5in,angle=-90}}
%\vskip 1cm
\caption{Phase and velocity relative to the wall at the moment of
reflection for $\alpha=165$ and $\eta=0.9999$, for three different
trajectories. To each trajectory, 50000 points are plotted.}
\label{transient}
\end{figure}
On the other hand, a stability analysis gives a stable fixed point for
\beq
\label{stability}
-{1+\eta\over 1-\eta}{2\over (2n-1)\pi} < \tan\phi<{1-\eta\over
1+\eta}{2\over (2n-1)\pi} \, .  
\eeq 
The violation of the left--hand
side of condition (\ref{stability}) corresponds to a period doubling
bifurcation (one eigenvalue becomes $-1$), the violation of the
right--hand side corresponds to a saddle--node bifurcation (one
eigenvalue becomes $+1$).

The chaotic regions surrounding the unstable fixed points for $\eta=1$
also continue to influence the dynamics for $\eta<1$. Typically
\cite{lie85}, chaotic regions become transient when a small dissipation
is added, and their main effect is the mixing of basins of
attraction. In figure \ref{transient} we illustrate this for
$\eta=0.9999$, where we show three different orbits of particles
starting at high initial velocities in the case $\alpha=165$. One
clearly sees the intertwining of different domains of attraction above
the first resonance, making the fate of each particle appear
random. After a long transient time, the first trajectory becomes
trapped in the first resonance, the third trajectory in the second
resonance, and the second trajectory is not yet trapped after 50000
reflections. It might either become attracted later by a
higher--period orbit, or it might be chattered regularly. In the
latter case, chatter should occur at intervals of the order of
$(1-\eta)^{-5/2} \simeq 3 \times 10^{12}$ reflections (see previous
subsection). 
A region that is chaotic for $\eta=1$ is not immediately transformed
into a strange attractor when $\eta$ is decreased \cite{lie85}.
However, strange attractors occur for smaller values of $\eta$, as we
shall see below.

\subsection{Large dissipation} \label{largediss}

As we have seen in Section \ref{inelastic}, every trajectory is
periodic for $\eta=0$ and may or may not go through the chattering
region $\phi \in [\pi,2\pi[ \hbox{ mod } 2\pi$. Both types of orbits
can coexist for small $\eta$ if the trajectory that starts in the
chattering region is not trapped by the other periodic orbit. Since
the basins of attraction of the chattering orbit and the normal
periodic orbit are strongly interwoven, a small change in $\eta$ can
induce or remove such a trapping. When $\eta$ is increased, normal
periodic orbits typically go through a period--doubling scenario and
then become a strange attractor. This attractor may be a global
attractor, or it may coexist with other attractors, e.g., periodic
orbits. Due to the large dissipation, the attractors are rather flat
and seem almost one--dimensional in Poincar\'e like plots, as also
known from other systems with large dissipation, like the Lorenz
attractor \cite{str94}. Fig.~\ref{attractor} shows a strange
attractor for $\alpha=10$ and $\eta=0.142$. 
The lower arches of the
attractor correspond to the second, third, and fourth reflection at
the same wall and are not present for smaller $\eta$. When $\eta$
increases further, the number of arches diverges, and the attractor is
replaced by a chattering orbit.

There are other mechanisms by which strange attractors may vanish with
increasing $\eta$. As the attractor increases with $\eta$, its edge
points may approach an unstable periodic orbit that ultimately diverts
the trajectory from the attractor and leads it either into the
chattering region, or to a stable periodic orbit that formerly
coexisted with the attractor. Alternatively, as in the Feigenbaum
scenario, an increase in $\eta$ can induce a transition from a strange
attractor to an intermittent periodic regime of some odd periodicity
(e.g. 3), that undergoes in turn a period--doubling scenario. Even in
cases where the attractor is destroyed due to chattering for some
value of $\eta$, such intermittent periodic regimes and subsequent
period--doubling cascades can still be observed when $\eta$ is further
increased.
\begin{figure}  \narrowtext
\centerline{\psfig{file=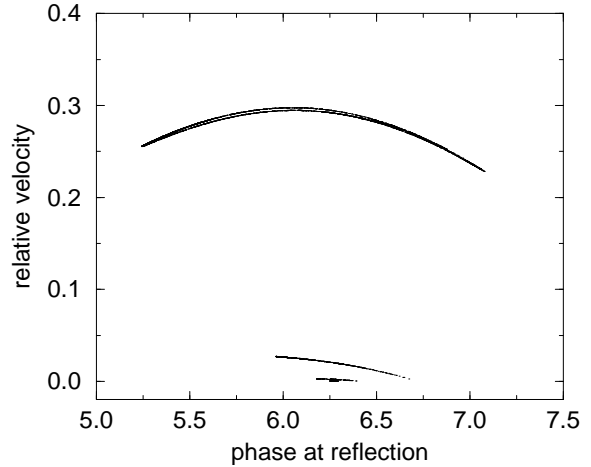,height=3in,angle=-90}}
%\vskip 1cm
\caption{The strange attractor for $\alpha=10$ and $\eta=0.142$.}
\label{attractor}
\end{figure}

Figure \ref{attractors} shows the various attractors for $\alpha=2$ as
$\eta$ is increased from 0 to 1.  Increasing $\eta$ from zero, we see
a stable fixed point emerging from the chattering domain which
dominates the dynamics at $\eta=0$.  This fixed point undergoes a
complete period doubling scenario as $\eta$ increases further. At the
end of the period doubling scenario we have a chaotic attractor which
continues to grow until it disappears when it overlaps with the
chattering domain. The latter part of the period doubling scenario
exists simultaneously with the stable three--cycle that is also shown
in the figure. This three--cycle moves to the left as $\eta$
increases, and it survives to $\eta=1$. Close to $\eta=1$, more
periodic orbits exist that are not shown in the figure.  Moreover,
there are other small attractors which appear for intermediate values
of $\eta$ (some of the ``dust'' in the figure).

We have chosen $\alpha=2$ for Fig.~\ref{attractors}, as here the
 period doubling is clearly visible. For other values of $\alpha$,
 more complicated scenarios can occur, as already indicated in the
 text above. There may be several period doubling cascades, which can
 occur for increasing as well as decreasing values of
 $\eta$. Period--doubling cascades for increasing $\eta$ are usually
 initiated by an eigenvalue of the stability matrix of a periodic
 orbit becoming equal to $-1$, while cascades for decreasing $\eta$
 occur when a periodic trajectory starts to experience several
 reflections at the same wall. Moreover, there can be intervals of
 $\eta$ for which all trajectories experience chattering, i.e., where
 the only attractor is the periodic orbit with
 chattering. Period--doubling cascades are also observed when $\alpha$
 is varied for fixed $\eta$.
\begin{figure}  \narrowtext
\centerline{\psfig{file=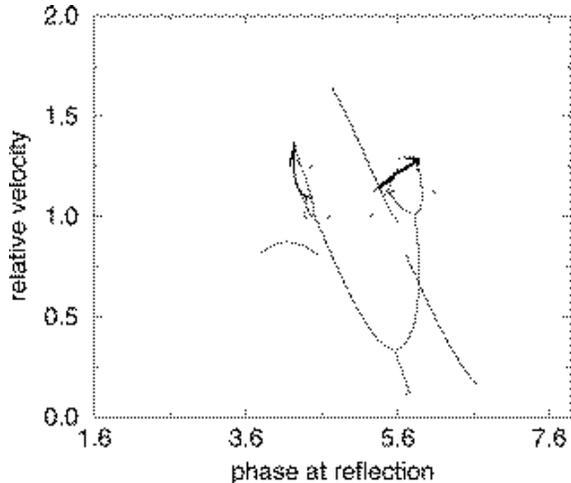,height=2.5in,angle=-90}}
%\vskip 1cm
\caption{The set of attractors for $\alpha=2$ as $\eta$ changes from
0 to 1.}
\label{attractors}
\end{figure}

\subsection{Ghost attractors and domains of attraction} \label{ghost}

Since the size of strange attractors increases with increasing $\eta$
until they become absorbed by the chattering orbit, strange attractors
become rare for higher values of $\eta$. Nevertheless, the ``ghosts''
of these attractors still continue to influence the dynamics by
trapping trajectories for a transient time, which can last several
hundred or thousand reflections. Consequently, the domains of
attraction for normal periodic and chattering orbits become strongly
interwoven, as shown in Fig.~\ref{domains} for $\alpha=10$ and
$\eta=0.61575$. 
There is a fixed point sitting in the center of
the curl, and there is one chattering orbit. The large black and white
areas around the curl are the domain of attraction of the fixed point,
and the two colours distinguish between trajectories that need an even or
odd number of reflections before reaching some small region around the
fixed point. The area with the irregularly distributed black dots is
the region where the domains of attraction of the fixed point and the
chattering orbit are interwoven, the black dots being points that
utimately become attracted to the fixed point. For two nearby starting
points in this region, the trajectories follow a ``ghost attractor'',
by which they are trapped for quite some time. The release of the
trajectories to either the attracting fixed point or the chattering
domain is seemingly random due to the influence of the attractor, as
can be seen from the bottom figure.  The white region at the bottom of
the top figure belongs entirely to the basin of attraction of the
chattering orbit.
\begin{figure}  \narrowtext
\centerline{\psfig{file=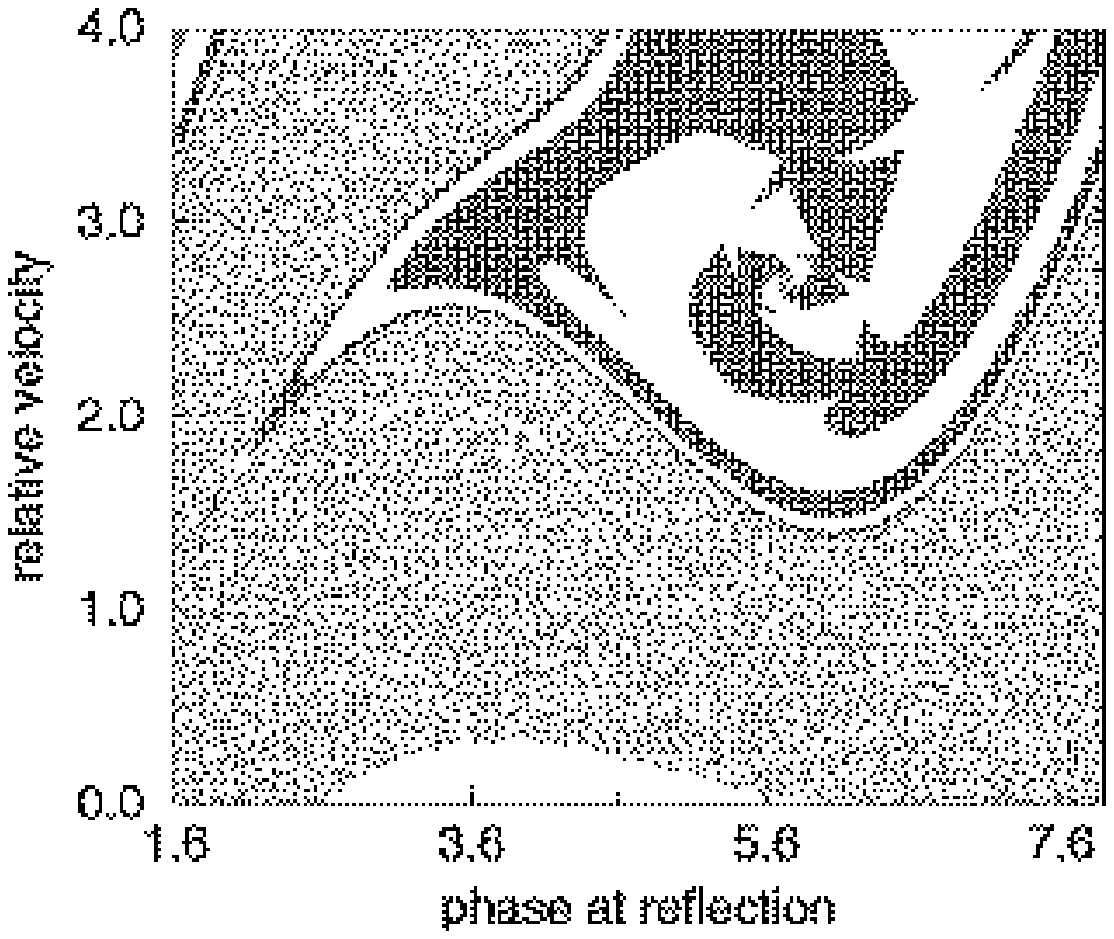,height=2.7in,angle=-90}}
\centerline{\psfig{file=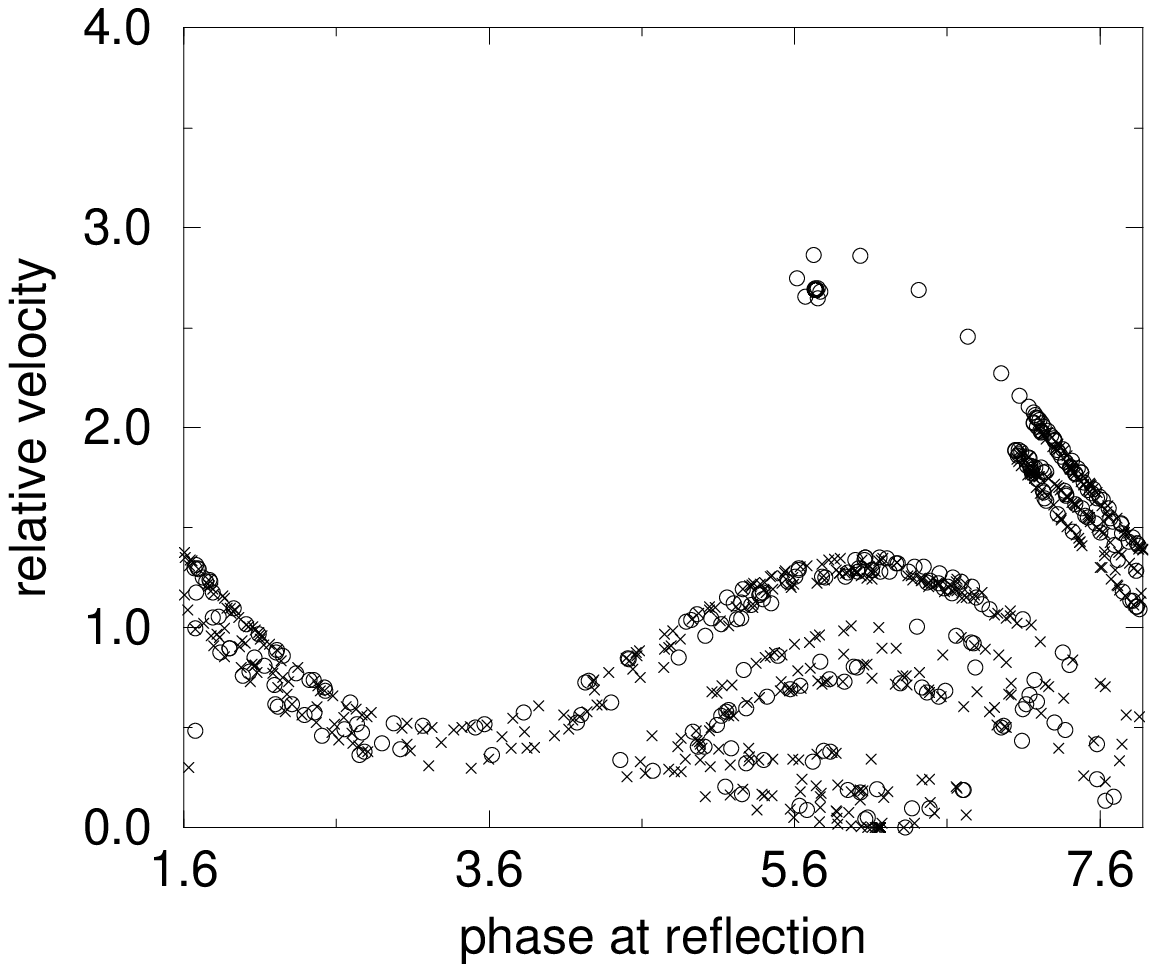,height=3in,angle=-90}}
%\vskip 1cm
\caption{The two domains of attraction for $\alpha=10$ and
$\eta=0.61578$.  The attracting fixed point is at $\phi\approx5.7$ and
$u\approx2.7$.  Chattering occurs immediately for orbits entering the
white domain at the bottom of the upper figure, and trajectories
starting in the large black and white areas around the fixed point
become absorbed in a small neighborhood around the fixed point after
an even resp.~odd number of steps. In the region with the irregularly
placed black dots, trajectories are strongly mixed, and it cannot be
predicted whether they will be chattered or absorbed at the fixed
point. The lower figure illustrates how how two nearby trajectories
follow a ``ghost attractor'', before they eventually either chatter or
get attracted to the fixed point.}
\label{domains}
\end{figure}

\subsection{The influence of friction} \label{friction2}
For the case of a completely inelastic particle, we have studied in
Section \ref{inelastic} the influence of friction between the particle
and the container bottom. Particles that are so slow that they cannot
reach the other wall become trapped in the region between the walls,
where they perform a low--amplitude periodic oscillation. For
sufficiently large container sizes, this is the fate of all particle
trajectories. A similar behaviour can be expected when the coefficient
of restitution does not vanish. All orbits that extend to a region of
sufficiently low velocity will become trapped in the region between
the walls. This will in particular affect long orbits like chattering
orbits and strange attractors.  For the completely elastic case, even
a small amount of friction destroys the hamiltonian nature of the
system, and we expect that the system behaves similar to the situation
where the coefficient of restitution is slightly smaller than 1. Thus, we
expect all stable periodic orbits to become attractive, and all
chaotic regions to become transient for an infinitesimally small
amount of friction.

\section{Conclusion and outlook}
\label{conclusion}
In this paper, we have shown that a particle in a horizontally shaken
box shows a much richer behaviour than a particle on a vibrating
platform. While chattering, i.e., the loss of all kinetic energy
during a finite time, may occur for any value of the coefficient of
restitution smaller than one, other scenarios like period--doubling
and strange attractors are observed as well. We have also discussed
the interplay and transitions between these scenarios, and the
influence of friction between the particle and the container bottom.
We know only of one other system where the interplay between chattering,
chaos, and periodic orbits has been studied before, namely the forced
oscillator impacting on a wall \cite{bud95}.

The chattering phenomenon discussed in this paper is a good
approximation to reality only when the collision time between the
particle and the wall is much shorter than the oscillation period. If
this condition is not satisfied, the deformation of the particle
during the collision has to be included in the model, thus adding
another degree of freedom. 

When $N>1$ particles are placed in the container, new phenomena will
arise. Three or more particles with a sufficiently small coefficient
of restitution $\eta < 1-1/N$ are known to undergo an {\it inelastic}
collapse, where they loose all their relative kinetic energy due to
infinitely many collisions during a finite time \cite{mcn92,zho96}, a
phenomenon similar to chatter. But even for parameter values that do
not allow for an inelastic collapse, clustering phenomena occur.  In
\cite{du95}, a system with one elastic and one thermally moving wall
is studied for approximately ten particles. The authors find that most
of the particles form a cluster almost at rest, while a few remaining
particles travel between the boundaries at a much higher speed. We
have seen a slightly different phenomenon in the periodically shaken
box, namely the formation of two clusters travelling between the
boundaries and the center of the system, similar to Newton's cradle.

\acknowledgements
This work was supported by EPSRC Grant No.~GR/K79307.

\end{multicols}
\end{document}